\documentclass[aps,pre,superscriptaddress,showpacs,twocolumn]{revtex4}

\usepackage{amsfonts}
\usepackage{amsmath}
\usepackage{multirow}
\usepackage{longtable}
\usepackage{graphicx}
\usepackage{color}

\newcommand{\be}{\begin{equation}}
\newcommand{\ee}{\end{equation}}
\newcommand{\bea}{\begin{eqnarray}}
\newcommand{\eea}{\end{eqnarray}}
\newcommand{\av}[1]{\langle #1 \rangle}

\begin{document}

\title{Random walks on activity-driven networks with attractiveness}

\author{Laura Alessandretti}
\email[]{l.alessandretti@gmail.com}
\affiliation{Department of Mathematics - City, University of London - Northampton Square, London EC1V 0HB, UK}

\author{Kaiyuan Sun}
\email[]{k.sun@neu.edu}
\affiliation{Laboratory for the Modelling of Biological and Socio-technical Systems,
Northeastern University, Boston MA 02115 USA}

\author{Andrea Baronchelli}
\email[]{a.baronchelli.work@gmail.com}
\affiliation{Department of Mathematics - City, University of London - Northampton Square, London EC1V 0HB, UK}

\author{Nicola Perra}
\email[]{n.perra@greenwich.ac.uk}
\affiliation{Centre for Business Network Analysis, University of Greenwich, Park Row, London SE10 9LS, United Kingdom}
\date{\today}

\begin{abstract}
Virtually all real-world networks are dynamical entities. In social networks, the propensity of nodes to engage in social interactions (activity) and their chances to be selected by active nodes (attractiveness) are heterogeneously distributed. Here, we present a time-varying network model where each node and the dynamical formation of ties are characterised by these two features. We study how these properties affect random walk processes unfolding on the network when the time scales describing the process and the network evolution are comparable. We derive analytical solutions for the stationary state and the mean first passage time of the process and we study cases informed by empirical observations of social networks. Our work shows that previously disregarded properties of real social systems such heterogeneous distributions of activity and attractiveness as well as the correlations between them, substantially affect the dynamical process unfolding on the network. 
\end{abstract}

\maketitle

Small-world phenomena along with heterogeneity in the number and frequency of contacts are among the most well known properties of social networks~\cite{newman10-1,jackson2008social,gonccalves2015social}. They are often referred to as late or time-integrated properties~\cite{holme11-1,holme2015modern} because they emerge integrating interactions over long time-scales. Traditionally, the modelling efforts put forward to characterise social systems and dynamical processes unfolding on their fabrics focused mainly on these features~\cite{newman10-1,barrat08-1}, neglecting the dynamics acting at much shorter time-scales. This was due to the challenges of introducing the temporal dimension in any mathematical construct and to the lack of real time-resolved datasets. While the former obstacle remains largely unsolved, significant progresses have been made to tackle the later~\cite{holme11-1,holme2015modern,masuda2016guide}. Indeed, the digital revolution has enabled scientists to access a wealth of offline and online data describing social interactions in time. The access to the temporal dimension allows to observe properties of social behaviour that are invisible in time-integrated datasets, and can help characterise microscopic mechanisms driving the dynamics of social acts at all time-scales~\cite{perra12-1,karsai13-1,ubaldi2015asymptotic,laurent2015calls,miritello2011dynamical,clauset07,Isella:2011,saramaki2015seconds,Saramaki21012014,Sekara06092016,tomasello2014role}. 
As a result, an intense research effort has been recently devoted to modeling the temporal dynamics characterising the emergence and evolution of networks. Furthermore, much attention has been directed to understand the effects of these dynamics on processes unfolding on the network such as the spreading of infectious diseases, idea, rumours, or memes~\cite{barrat2015face,perra12-2,perra12-1,ribeiro12-2,PhysRevLett.112.118702,PhysRevE.87.032805,10.1063,starnini13-1,starnini_rw_temp_nets,valdano2015analytical,scholtes2014causality,Williams160196,rocha2014random,takaguchi2012importance,rocha2013bursts,ghoshal2006attractiveness,sun2015contrasting,pfitzner2013betweenness,takaguchi12-1,takaguchi2013bursty,holme2014birth,holme2015basic,dynnetkaski2011,hoffmann2012generalized,gonccalves2015social,wang2016statistical,fournet2014contact}.

Observations in a range of real social networks show that the propensity of individuals to engage in social acts is highly heterogenous~\cite{perra12-1,ribeiro12-2,karsai13-1,ubaldi2015asymptotic,tomasello2014role}. Also, it was found that the establishment of connections is highly correlated in time~\cite{Karsai2011Small,moinet2015burstiness,karsai12-1,karsai13-1,ubaldi2015asymptotic,laurent2015calls}. Several studies have focused on understanding the effects of \emph{local} memory in the creation of links. It was shown that different types of local reinforcement mechanisms are able to mimic characteristic aspects of social networks such as the emergence of strong and weak ties~\cite{karsai13-1,ubaldi2015asymptotic,ubaldi2016burstiness,laurent2015calls,Onnela:2007,granovetter73-1}.   

However, in certain circumstances \emph{local} mechanisms alone can not explain the creation of social ties. For example, in online social networks like Twitter individuals can interact with popular figures and access topical pieces of information. Arguably, the creation of these connections does not follow the same local rules driving the emergence of close social ties. Instead, at least to some extent, they may be driven by \emph{global} effects such as interest towards celebrities or for the information provided by popular accounts. Despite the widespread diffusion of these platforms, the modelling of global mechanisms for link creation and the understanding of its effects on diffusion processes unfolding on the network remain largely unexplored. This is especially true when short-time scales and thus time-varying dynamics are considered.

In this paper, we propose a temporal model of interactions driven by global popularity. In particular, we extend the activity-driven framework~\cite{perra12-1} in which individuals/nodes are assigned an activity defining their propensity to establish contacts per unit time. In its first formulation active nodes connect to others through a memoryless and random selection process~\cite{perra12-1}. More realistic mechanisms based on local reinforcement of ties have been then proposed~\cite{karsai13-1,ubaldi2015asymptotic,ubaldi2016burstiness}. Here, we present a new variation in which nodes are characterised by an attractiveness~\cite{PhysRevE.87.062807, starnini2013modeling,starnini2016model,mariani2015ranking}, or a popularity index, that might or might not be correlated with activity and drive the contact selection process. In particular, we consider a classic linear preferential attachment~\cite{barabasi2016network}. We then study a random walk process unfolding at the same time-scale in which the connections are created. For sake of simplicity, we consider the fundamental random walk process, which has recently been investigated on different kinds of temporal networks~\cite{perra12-2,ribeiro12-2,rocha2014random,scholtes2014causality,hoffmann2012generalized,lambiotte2013burstiness,lambiotte2015effect}. We find analytical solutions for the stationary state of the process as well as its mean first passage time (MFPT) that match the results produced by numerical simulations. The solutions are general and allow to analytically characterise the interplay between activity and attractiveness considering also their correlations.
We ground our results with empirical observations by measuring such correlations on different real datasets and we discuss their repercussions on the random walks. 

The paper is organised as follows. In Section~\ref{model} we introduce the network model.  In Section~\ref{correlation}, we study the interplay between activity and attractiveness in real networks. In Section~\ref{RW} we study the stationary state of the random walks diffusing on the model. In Section~\ref{mfpt} we study the MFPT. Finally, in Section~\ref{conclu} we discuss our conclusions. 

\section{Time-varying network model}
\label{model}

\begin{figure}
\includegraphics[width=.5\textwidth]{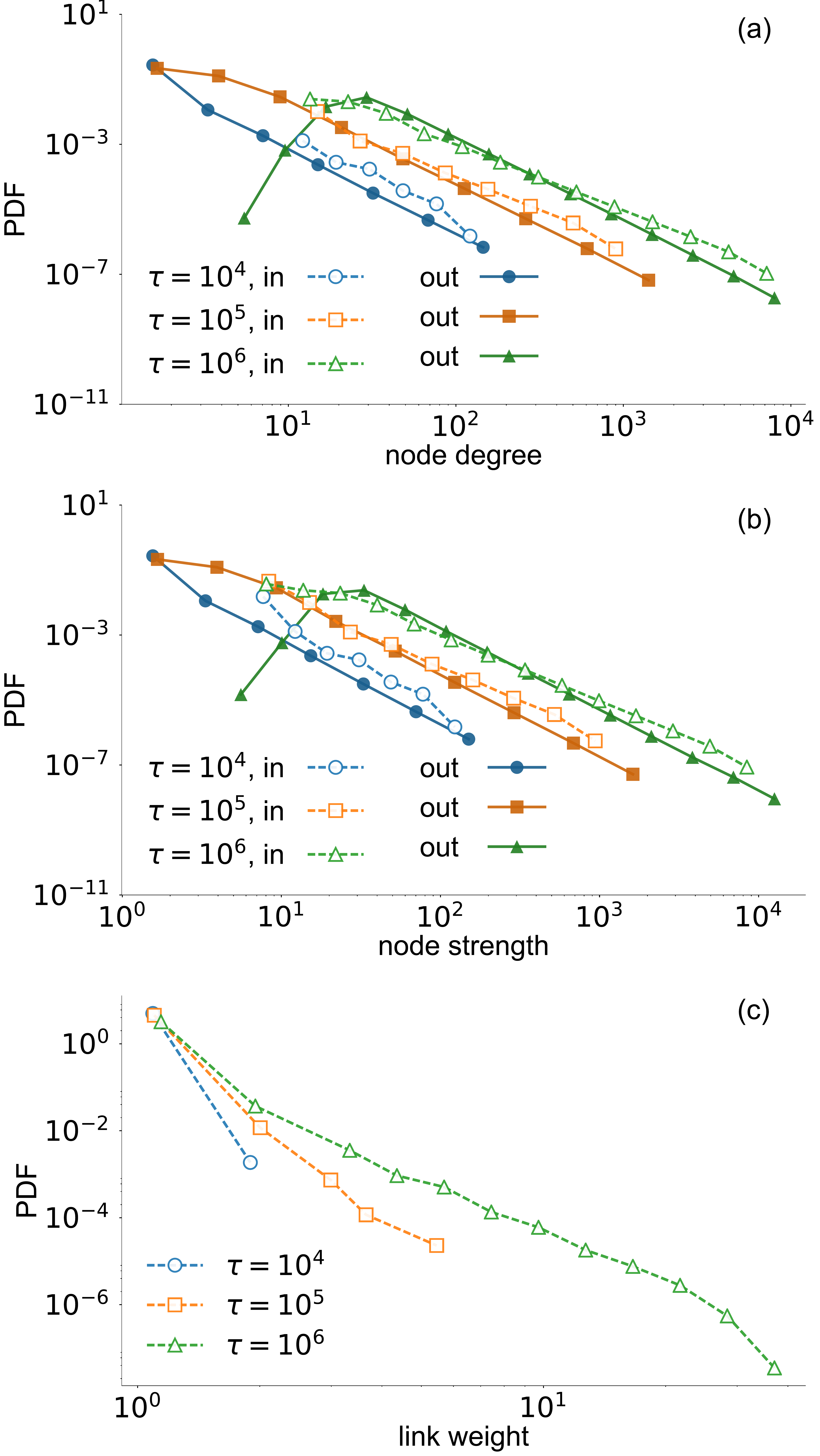}
\caption{\textbf{Statistical properties of the time-aggregated network} Probability density function of nodes of given in and out-degree (a) and strength (b) for different values of time-window $\tau$. Probability density function of links of given weight for different values of the time-window $\tau$ (c). Results are shown for $N=10^5$, $m=6$, $F(a)\sim a^{-2}$, $G(b)\sim b^{-2.5}$. For $\tau= 10^4$, the average in-degree is $\langle k_{in} \rangle = 0.6$, for $\tau= 10^5$, $\langle k_{in} \rangle = 5.7$, for $\tau= 10^6$, $\langle k_{in} \rangle = 57.9$. Note that the average out-degree equals the average in-degree.}
\label{evolution}
\end{figure}

In the activity-driven network framework~\cite{perra12-1} the $N$ nodes of the network are assigned an activity rate $a$ describing their propensity to engage in social acts~\cite{perra12-1,ribeiro12-2,karsai13-1,tomasello2014role,ubaldi2015asymptotic}. Here, we consider nodes characterised also by another quantity, namely their \textit{attractiveness} $b$, describing their popularity in the system~\cite{starnini2013modeling,starnini2016model,starnini2016emergence}. In general, these two quantities are correlated and extracted from a joint distribution $H(a,b)$. 
At each time step, a node $i$ is activated with probability $a_i \Delta t$ and connects to $m$ others.  The generic node $j$ is selected with probability $b_j/\av{b}N$. Each link has a duration of $\Delta t$. In Figure \ref{evolution}, we show the statistical features of the emerging network considering $N=10^5,m=6$ for an uncorrelated system where$H(a,b)= F(a)G(b) \sim a^{-2} b^{-2.5}$, integrating over time $\tau$. Here, $\tau$ is expressed in units of the average time between consecutive activations $a_0^{-1}$, where $a_0 = \sum_i a_i=\langle a \rangle N$\cite{gillespie1977exact}. As clear from the figure, the heterogeneity in activity and attractiveness induces heavy-tailed degree, strength, weight distributions. This is analogous to what is observed in the case of nodes with heterogeneous activity. \\

\section{Correlation between activity and attractiveness in real networks}
\label{correlation}
\begin{figure}[t!]
\includegraphics[width=0.5\textwidth]{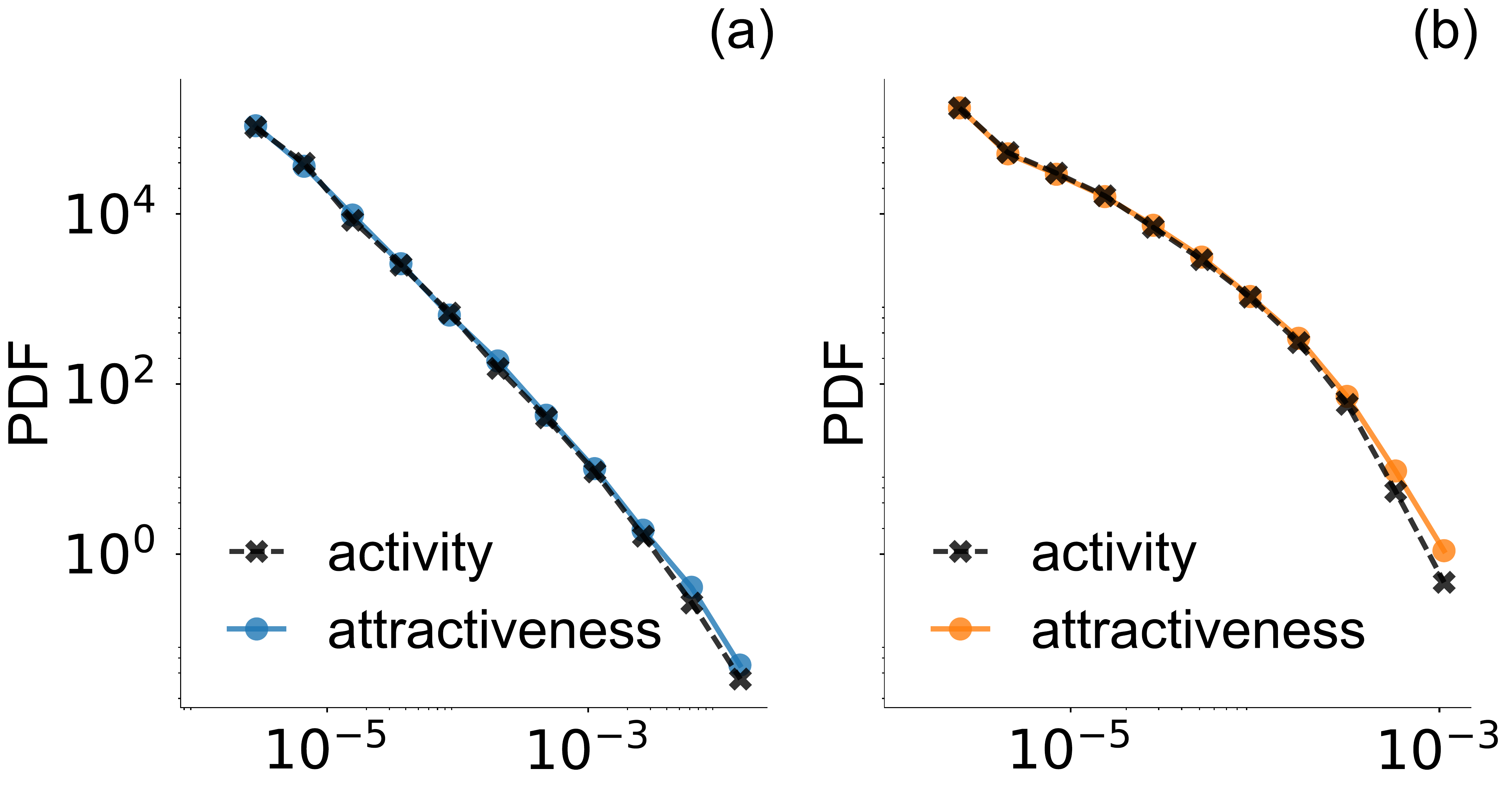}
\caption{\textbf{Distribution of activity and attractiveness in real datasets.} Probability density function of activity ($a$) and attractiveness ($b$) for the Linux dataset (a) and the Facebook dataset (b). In the Linux network there are $N=2.7 \cdot 10^4$ nodes, $E = 1.0 \cdot 10^6$ edges, and the period of measurement lasts $T=2921$ days. For the Facebook network, $N=4.6 \cdot 10^4$, $E= 8.6 \cdot 10^5$, $T=1591$ days }
\label{data}
\end{figure}

\begin{figure}[t!]
\includegraphics[width=0.5\textwidth]{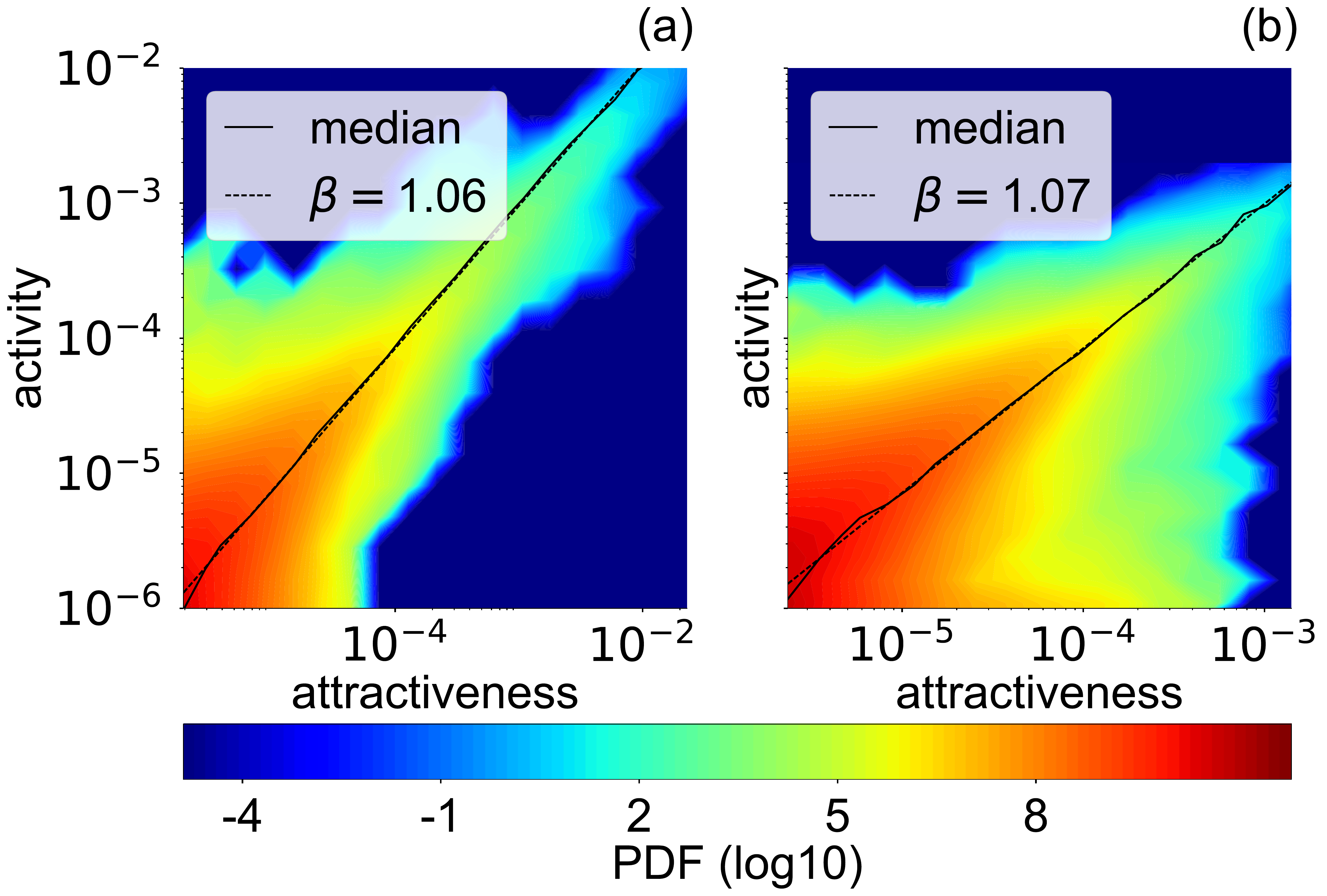}
\caption{\textbf{Correlation between activity and attractiveness in real datasets.} Heat map showing the correlation between activity and attractiveness in two real datasets describing interactions between people involved in the development of Linux (a) and on Facebook (b). The continuous line describes the median correlation and the dashed line  a power-law fit with exponent $\beta$. }
\label{correlation_a_b}
\end{figure}

The activity measures the propensity of nodes to initiate a social interaction, while attractiveness quantifies the probability of being selected to participate to such interactions, i.e. popularity. These two quantities and their correlation can be studied in real networks, provided that interactions are directed and allow to distinguish between the activation and selection process. Here, we consider two datasets. The first describes wall-posts interactions between $45,813$ Facebook users over a timespan of $1,591$ days \cite{viswanath2009evolution,konect:2016:facebook-wosn-wall}. The second describes email replies among $26,885$ users involved in the Linux kernel development over $2,921$ days \cite{konect:2016:lkml-reply}. For the sake of this model, we consider the out-strength and in-strength of nodes as proxies for their activity and attractiveness respectively. Hence, activity and attractiveness of node $i$ are computed as $a_i = s_{i,out}/\sum_j s_{j,out} $ and $b_i = s_{i,in}/\sum_j (s_{j,in}) $, where $s_{i,in}$ and $s_{i,out}$ are the node in-strength and out-strength integrated across the entire time-span, respectively. Activity and attractiveness are computed aggregating across the whole period of data collection. In fact, observations in a range of real datasets such as co-authorship networks~\cite{perra12-1,ubaldi2015asymptotic}, online social networks~\cite{perra12-1,ribeiro12-2} mobile phone networks~\cite{karsai13-1}, and networks created by R\&D alliances between firms~\cite{tomasello2014role} show that the form of the activity distribution is independent of the aggregation window. In Figure \ref{data} we show the distributions of activity and attractiveness in the two datasets. Not surprisingly, in the two datasets both activity and attractiveness follow heavy-tailed distributions spanning several order of magnitude \cite{barrat2004architecture}. In Figure~\ref{correlation_a_b} we plot the correlation between activity and attractiveness considering each node in the two datasets. A positive correlation is clear and in both cases the median follows a power-law with exponent very close to one, i.e. $a\sim b^{\beta}$, $\beta \sim 1$.

\section{Random walk}
\label{RW}

We consider a Markovian and homogenous random walk~\cite{noh04} unfolding on networks generated with the model described above. We focus on the case in which the walker moves at the same time scale describing the evolution of links, moving from node to node when a link is present. The properties of the diffusion process thus are highly affected by the dynamics driving the evolution of the connections. \\

Let us define $P_i(t)$ as the probability that the walker is in node $i$ at time $t$. This quantity follows the following master equation:

\be
\label{me1}
P(i,t+\Delta t)= P(i, t)[1- \sum_{j \neq i} \Pi_{i \rightarrow j}^{\Delta t}]+ \sum_{j \neq i}P(j,t)\Pi_{j \rightarrow i}^{\Delta t},
\ee 
where $ \Pi_{i \rightarrow j}^{\Delta t}$ is the propagator of the random walk that describes the probability that the walker moves from $i$ to $j$ in a time interval $\Delta t$. A link between $i$ and $j$ can be created as consequence of the activation of $i$ or $j$. The probability that $i$ is active and selects $j$ is:
\be
p(i\rightarrow j)=\frac{m a_i\Delta{t} b_j}{N\langle b\rangle}.
\ee
In this case, the instantaneous degree of $i$ is:
\be
k_i=m + \frac{m \langle a \rangle \Delta t b_i}{N\langle b\rangle}.
\ee
Indeed, $i$ will generate $m$ links and will potentially receive links from other active nodes. The probability that $j$ is active and selects $i$ is instead:
\be
p(j\rightarrow i)=\frac{m a_j\Delta{t} b_i}{N\langle b\rangle}.
\ee 
The instantaneous degree of $i$ will be:
\be
k_i=1 + \frac{m \langle a \rangle \Delta t b_i}{N\langle b\rangle}.
\ee
In the limit $\Delta t \rightarrow 0$, the events described by equations (2) and (4) do not happen simultaneously. 
Putting all together is easy to show that, for $\Delta t \rightarrow 0$:

\bea
\label{prop}
\Pi_{i \rightarrow j}^{\Delta t}&=&\frac{m a_i\Delta{t} b_j}{N\langle b \rangle}\frac{1}{m + \frac{m \langle a \rangle \Delta t b_i}{N\langle b \rangle} }+\frac{ma_j\Delta{t}b_i }{N\langle b \rangle}\frac{1}{1+\frac{m \langle a \rangle \Delta t b_i}{N\langle b \rangle}} \nonumber \\
&\simeq& \frac{\Delta t}{N\langle b \rangle} (a_i b_j+m a_j b_i).
\eea
\begin{widetext}
In the limit $\Delta t \rightarrow 0$ we can write the equation describing the evolution of $P_i(t)$ by substituting the expression of the propagator in Eq.~\ref{me1}:
\bea
\label{me2}
\dot{P}(i,t) = -\frac{P(i,t)}{N\langle b \rangle}\sum_{j \neq i} (a_ib_j+ma_jb_i)+\sum_{j \neq i}\frac{P(j,t)}{N\langle b \rangle}(a_jb_i+ma_ib_j)=\nonumber \\
-\frac{P(i,t)}{\langle b \rangle}[a_i  \langle b \rangle + m b_i  \langle a \rangle ]+ \frac{b_i}{N\langle b \rangle} \sum_jP(j,t)a_j+ \frac{ma_i}{N\langle b \rangle} \sum_jP(j,t)b_j.
\eea
\end{widetext}

\begin{figure}
\includegraphics[width=0.5\textwidth]{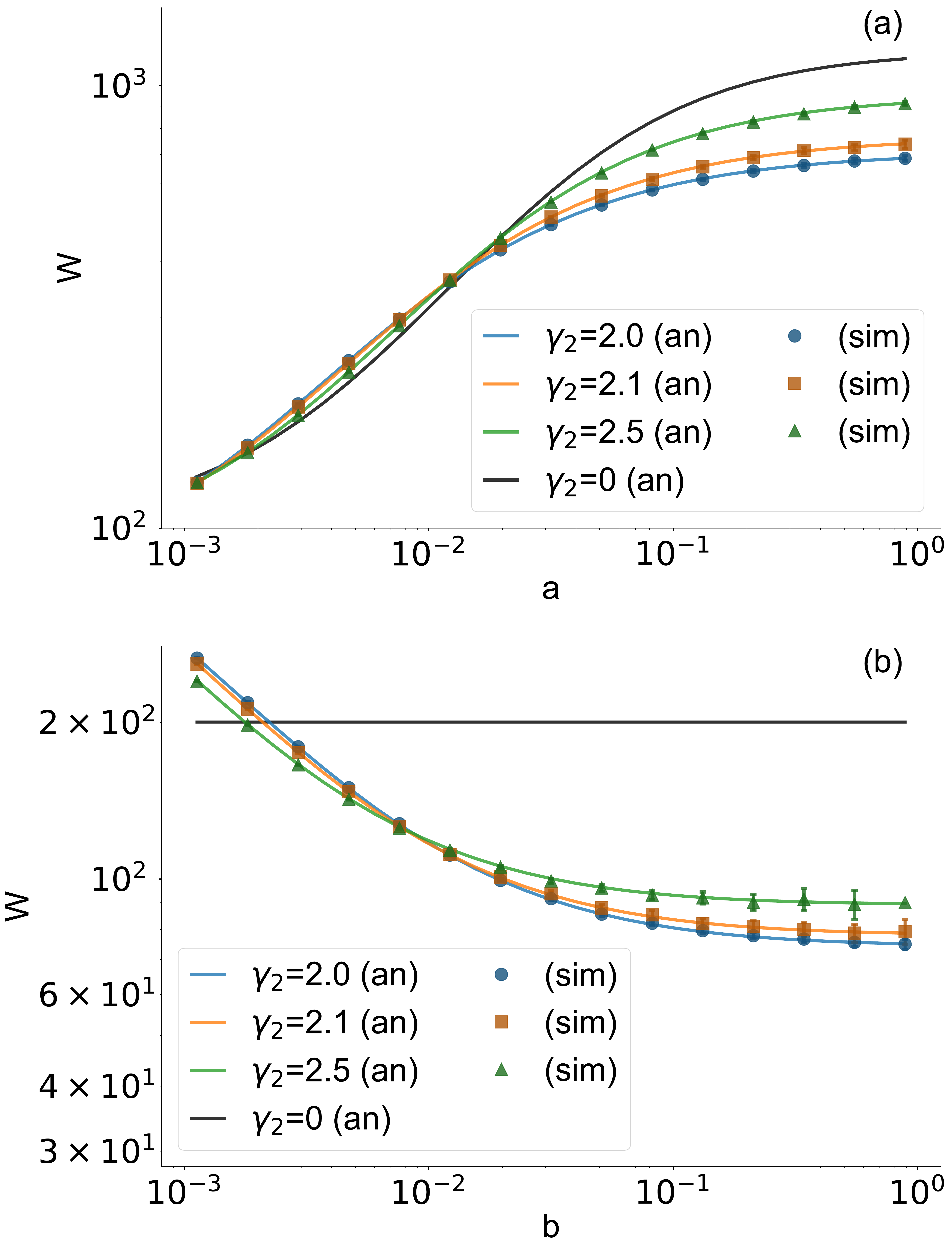}
\caption{\textbf{Stationary state of the random walk process.} The average number of walkers per node of class $a$  (a) and $b$ (b) computed analytically (continuous lines) and through numerical simulations (dots, squares and triangles) for different values of exponent of the attractiveness distribution $\gamma_2$. Error-bars are standard deviations obtained by averaging across $10^3$ different network configurations. In the above panel, they are not visible on the scale of the graph. We considered $N=10^5$, $m=6$, $W/N=200$, and $\gamma_1=2$. The black line shows the case where $b_i=1/N$ for all nodes. }
\label{same_1d}
\end{figure}

\begin{figure}
\includegraphics[width=0.5\textwidth]{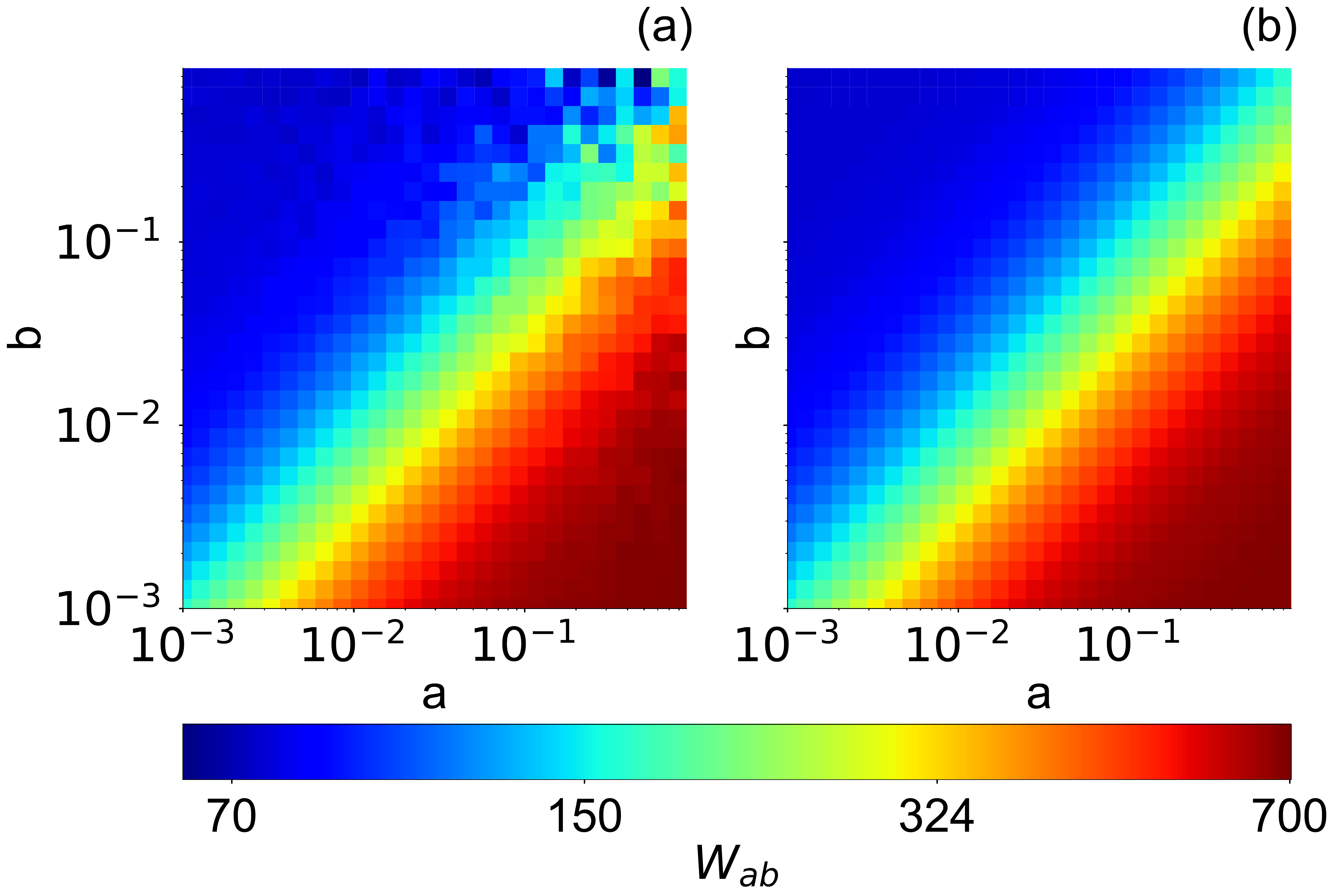}
\caption{\textbf{Stationary state of the random walk process for nodes of class $(a,b)$.} Heat map giving the average number of walkers $W_{ab}$ per node of class $(a,b)$ computed through numerical simulations (a) and analytically (b). Colors are attributed based on $W_{ab}$ as shown in the colorbar on the bottom of the figure. We considered $N=10^5$, $m=6$, $W/N=200$, $\gamma_1=2$, and $\gamma_2=2$.}
\label{same_2d}
\end{figure}

We obtain a system level description of the process by grouping nodes in the same activity class $a$ and attractiveness $b$, assuming
that they are statistically equivalent~\cite{alex12-1}. Then, we define the walkers in a given node of class $a$ and $b$ at time $t$ as
$W_{ab}(t) = [NH (a,b)]^{-1}W \sum_{i \in a \& \in b} P_i(t)$, where, $W$ is the total number of walkers in the system. By considering the continuous $a$ and $b$ limit, Eq.~\ref{me2} can be rewritten as:
\bea
\dot{W}_{ab}(t)&=& -\frac{W_{ab}(t)}{\langle b \rangle}[a \langle b \rangle + m b  \langle a \rangle] \nonumber \\
&+& \frac{b}{ \langle b \rangle} \iint a'W_{a'b'}(t)H(a',b') da'db' \nonumber \\
&+& \frac{ma}{\langle b \rangle}\iint b'W_{a'b'}(t)H(a',b') da'db' \\=
&-&\frac{W_{ab}(t)}{\langle b \rangle}[a \langle b \rangle + m b \langle a \rangle]+
\frac{b}{ \langle b \rangle}\phi_1+\frac{ma}{ \langle b \rangle}\phi_2,
\eea
where $\phi_1=\iint a'W_{a'b'}(t)H(a',b') da'db' $ and $\phi_2=\iint b'W_{a'b'}(t)H(a',b') da'db'$. In the stationary state, the changes of $W_{ab}(t)$ are zero, thus we have:
\be
W_{ab}(t)=  \frac{b\phi_1+ma\phi_2}{a\langle b \rangle +mb\langle a \rangle}.
\ee
The stationary state features $a$ and $b$ in both numerator and denominator. Hence, the dynamical properties of the random walk are function of the interplay between the two quantities. It is important to notice that at the stationary state $\phi_1$ and $\phi_2$ are constant. Their value can be computed self-consistently by solving this system of integral equations:
\bea
W&=& N \iint H(a,b)\frac{b\phi_1
+ma\phi_2}{a\langle b \rangle+mb\langle a \rangle}dadb  \nonumber \\
\phi_2&=&\iint bH(a,b)\frac{b\phi_1
+ma\phi_2}{a\langle b \rangle+mb\langle a \rangle}dadb,
\eea
where the first equation follows from the conservation of walkers in the system.\\
We test the analytical solutions against numerical simulations run following the Gillespie algorithm \cite{gillespie1977exact}. As a first step, let us consider the uncorrelated case in which both $a$ and $b$ are extracted from a power-law distribution: $H(a,b)=F(a)G(b)$ where $F(a)=Aa^{-\gamma_1}$ and $G(b)=Cb^{-\gamma_2}$. In both cases values are extracted in the range $x \in [10^{-3}, 1]$. In Figure~\ref{same_1d} we plot the comparison between the average number of walkers per nodes of class $a$ and $b$ separately. In Figure~\ref{same_2d} we plot instead $W_{a,b}(t)$ as a heat map. In both cases, the agreement between simulations and analytical predictions is clear. \\

Taken together, the two figures present a rich picture. First, they show that the larger the activity, the larger the capability of gathering walkers. The trend holds up to a saturation point after which an increase in activity does not translate to an increase of walkers, similarly to what is observed in Ref.~\cite{perra12-2} for the case of constant attractiveness, i.e. random tie selection process (see also Figure~\ref{same_1d}, bottom panel, black filled line). Second, they reveal an opposite trend for increasing values of $b$, as, before saturation, the larger the attractiveness the smaller the number of walkers in the stationary state. While this finding could seem counterintuitive, it can be understood considering the structure of the instantaneous network where walkers move. In the limit $\Delta t \rightarrow 0$, the degree of an active node $i$ is $k_i \sim m$, while the degree of a node $j$ connected by $i$ is $k_j \sim 1$ as non-active nodes do not `have time to' accumulate multiple connections. Thus, even extremely attractive nodes, that are involved in many connections across time, appear instantaneously as nodes with degree $1$. Consequently, a node selected by $i$ receives on average a fraction $1/m$ of the walkers of $i$, but it sends all its walkers to $i$. This fact explains the decreasing trend of $W(b)$ and shows at a fundamental level the effects of temporal interactions on diffusion processes taking place on the same timescale. As a consequence, in the case of a random-tie selection process nodes with large activity are able to collect more walkers than in the case of heterogeneous $b$, due to the tendency to select nodes holding fewer walkers than average in the latter case. 

\begin{figure}
\includegraphics[width=0.5\textwidth]{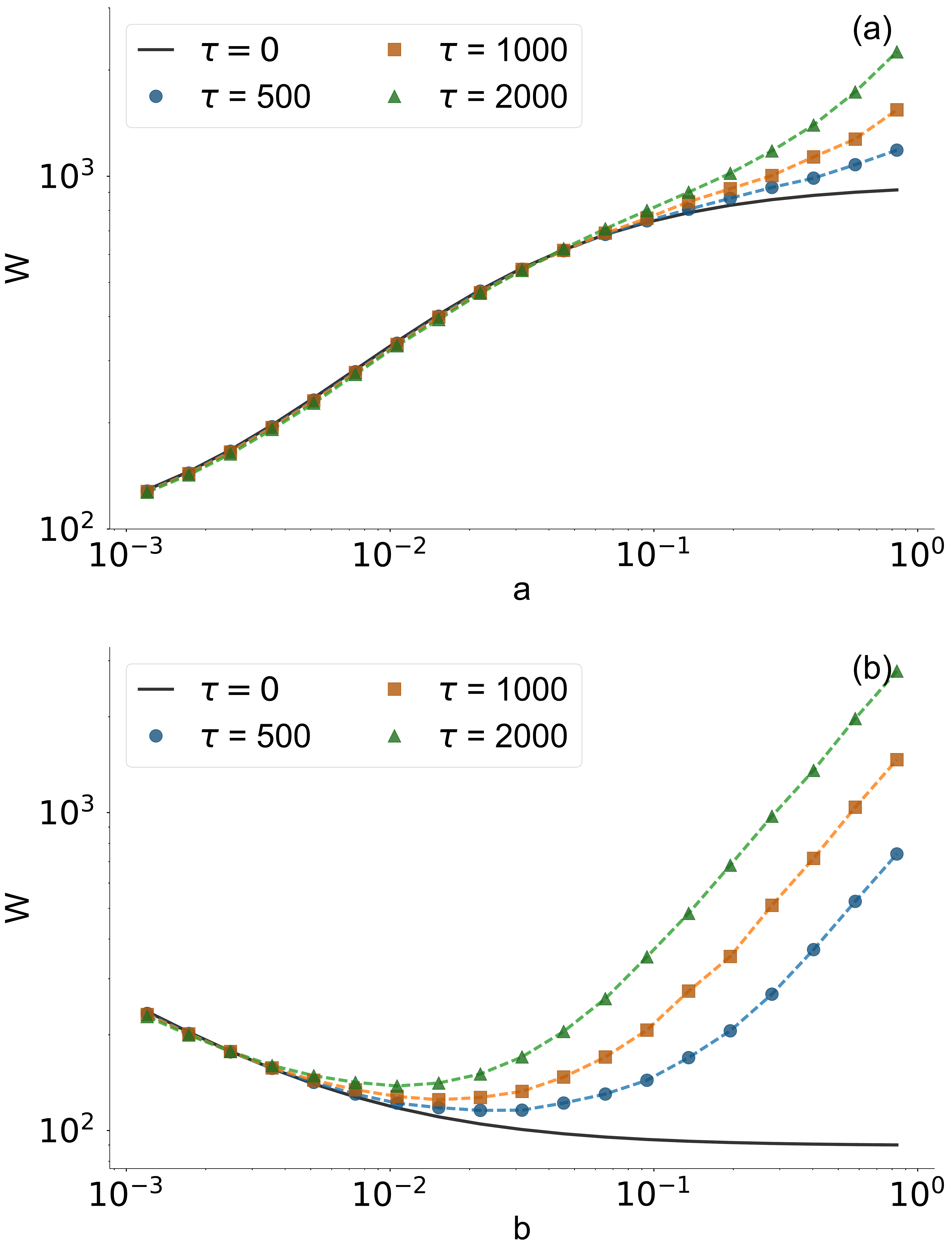}
\caption{\textbf{Stationary state of the random walk in the aggregated case.} The average number of walkers per node of class $a$  (a) and $b$ (b) computed analytically for $\tau=0$ (continuous line) and through numerical simulations for several values of $\tau$ (dots, squares and triangles). Dashed lines are shown as a guide for the eye. Error-bars obtained by averaging across $10^3$ different network configurations are not visible on the scale of the graph. We considered $N=10^5$, $m=6$, $W/N=200$, $\gamma_1=2$, and $\gamma_2=2.5$. }
\label{aggregated_time}
\end{figure}

To further understand these effects, we study the case of random walks unfolding on static networks obtained by integrating activity-driven networks with attractiveness over time windows of size $\tau$. In doing so, we let nodes activate and connect to other nodes for a time $\tau$. Then, we let the random walk unfolds on the union of such networks. Note that, in this case, interactions are not instantaneous. In figure \ref{aggregated_time}, we show the stationary state of the process as a function of the nodes activity and attractiveness, for different value of $\tau$. In contrast to what observed when the diffusion process and the topology evolve at the same timescale, here the walkers concentrate also on highly attractive nodes. This result is expected. The stationary state of random walks unfolding on any static network is linearly proportional to the degree~\cite{noh04,newman10-1}. In our case, nodes with large attractiveness are likely to be hubs: characterised by large degree values. These effects are more evident for increasing values of the time-aggregation window. Indeed, the larger $\tau$, the larger the degree of highly attractive nodes. For similar reasons, the same qualitative behaviour is observed also for nodes with high activity.

\begin{figure}
\includegraphics[width=0.5\textwidth]{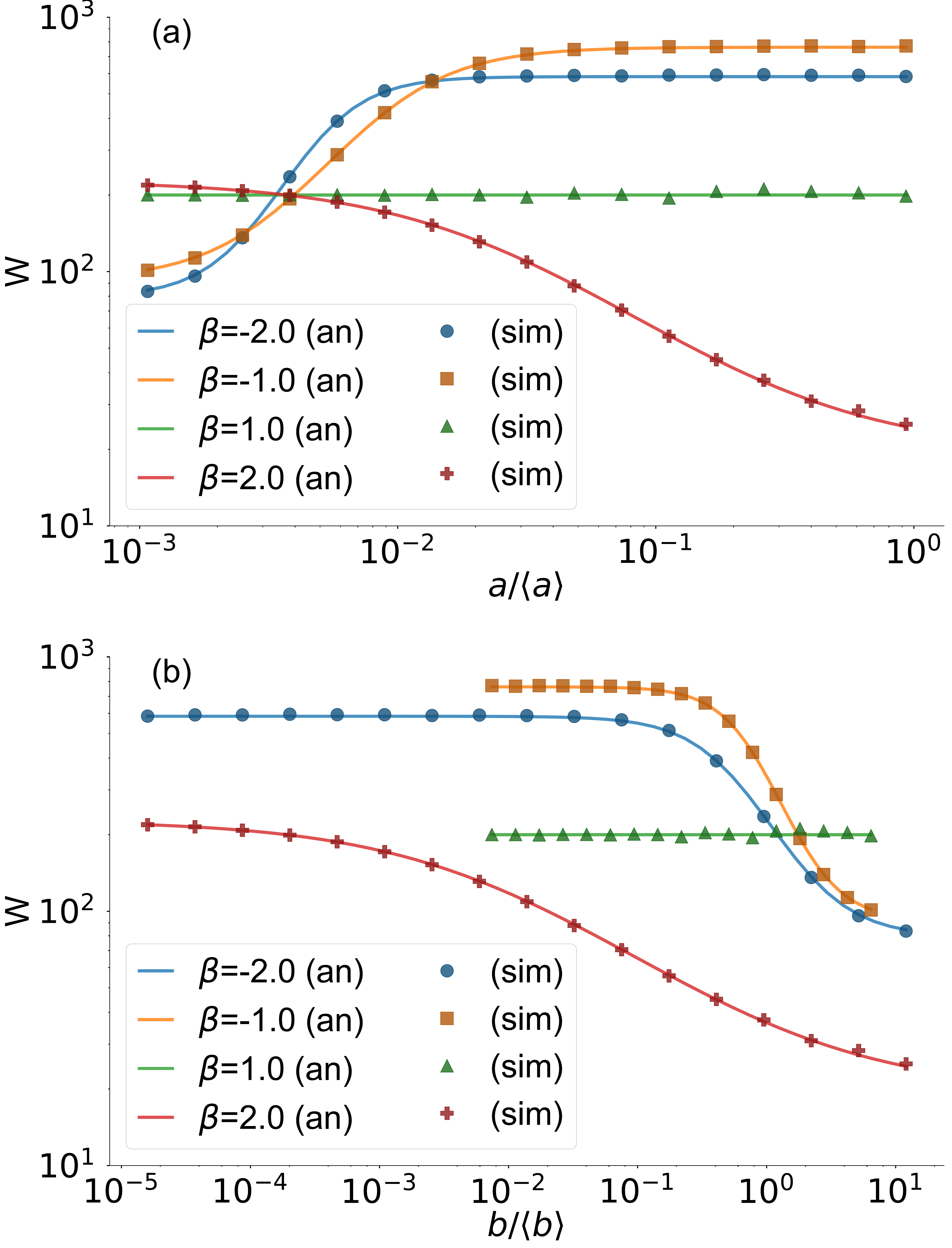}
\caption{\textbf{Stationary state of the random walk in the correlated case.} The average number of walkers per node of class $a$  (a) and $b$ (b) computed analytically (continuous lines) and through numerical simulations (dots, squares, triangles and crosses) for different values of the correlation exponent $\beta$. Error-bars obtained by averaging across $10^3$ different network configurations are not visible on the scale of the graph. We considered $N=10^5$, $m=6$, $W/N=200$, and $\gamma_1=2$. }
\label{correlations}
\end{figure}

Considering the observations in real datasets, we turn now the attention to scenarios in which activity and attractiveness are correlated. In particular, we consider for each node a deterministic correlation of the form $b \sim a^{\beta}$, or more in general $b=J(a)$ where $J$ is a generic function. The joint probability can then be written as $H(a,b)=F(a)\delta(b-J(a))$, where $\delta(x)$ is the Dirac delta. In Figure \ref{correlations} we show the stationary state of the random walks for several values of $\beta$.

For $\beta<1$ trends are not far from the uncorrelated case. For larger activity, nodes have higher capability of gathering walker and $W(a)$ saturates for large values of $a$, while the opposite trend holds for $W(b)$. Indeed, the negative correlation reinforces what is observed in the uncorrelated case since nodes with low-activity have also high attractiveness. Hence, we observe that the larger $\beta$, the smaller is the number of walkers collected by nodes with low activity and the faster $W(a)$ saturates. 

Instead, for $\beta>1$, the larger the activity, the lower the capability of gathering walkers. In this case, the set of nodes more frequently engaged in active interaction has also attractiveness much larger than the average node. These nodes tend not to hold walkers but to exchange them continuously. Instead, walkers are likely to be trapped in nodes that are unlikely to engage in interaction.

For $\beta=1$, since the rate at which node is activated and the probability to be selected are exactly the same, $W(a)$ and $W(b)$ are constant.

\subsection{MFPT}
\label{mfpt}

We now consider the mean first passage time (MFPT), defined as the average number of time steps needed for a walker to visit a node $i$ starting from any other node in the system~\cite{Redner01,noh04,baronchelli2006ring}. 

Let us consider $p(i,n)$ as the probability that the walker reaches $i$ (the target) for the first time at time 
$t=n\Delta t$. Considering that each node could be connected directly to any other, we have:
\be
p(i,n)=\xi_i (1-\xi_i)^{n-1},
\ee
where $\xi_i$ is the probability that the walker jumps in node $i$ in a time interval $\Delta t$, that is:
\be
\xi_i=\sum_j \frac{W(a_j,b_j)}{W}\Pi^{\Delta t}_{j \rightarrow i}.
\ee
Indeed, the propagator by definition encodes the probability that walkers moves from $j$ to $i$, and $W(a_j,b_j)/W$ describes the probability that the walker is in $j$ at time $t$ (in the stationary state). Thus, we can estimate the MFPT as:
\bea
\label{mfpt_exp}
MFPT_i &=& \sum_{n=0}^{\infty}  n\Delta t \cdot p(i,n)=\frac{\Delta t}{\xi_i} \nonumber \\
&=& \frac{N\langle b \rangle w}{b_i \sum_j W(a_jb_j) a_j+m a_i \sum_j W(a_jb_j) b_j} \nonumber \\
&=& \frac{\langle b \rangle w}{b_i \phi_1+m a_i \phi_2}. 
\eea
It is interesting to notice how in static and annealed networks (where the timescale of the random walk is either much faster or slower with respect to changes in the topology where it is unfolding) $\xi_i$ is equivalent to the stationary state of the random walk, i.e. $\xi_i = W_i/W$. In time-varying networks instead this is not the case as the walker can be trapped in an inactive or unpopular node for several time steps~\cite{perra12-2}. Consequently, the expression of $\xi$ considers explicitly the dynamical connectivity patterns to account for such delays. 

\begin{figure}
\includegraphics[width=0.5\textwidth]{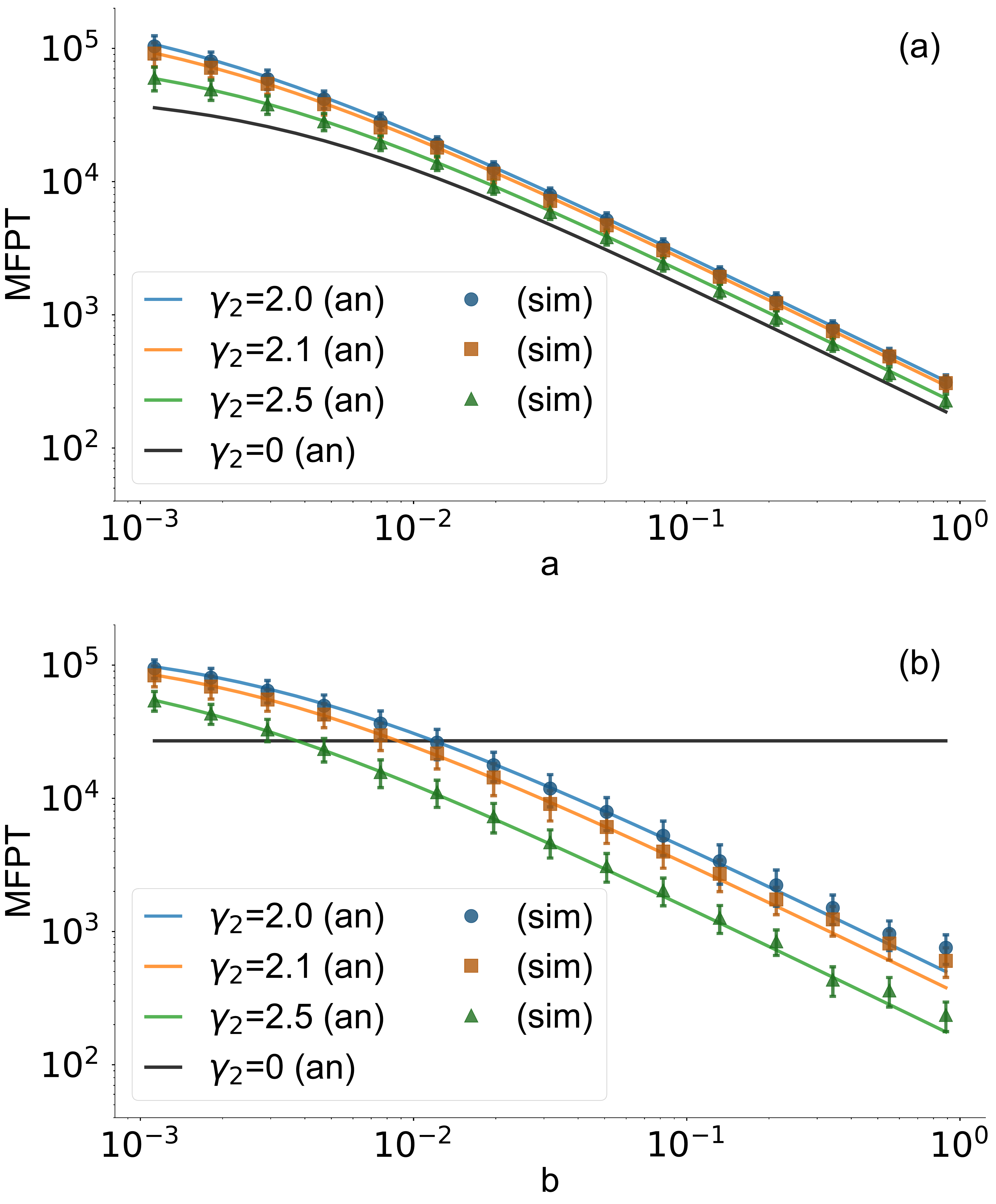}
\caption{\textbf{Mean first passage time.} The average MFPT as function of $a$  (a) and $b$ (b) computed analytically (continuous lines) and through numerical simulations (dots, squares and triangles) for different values of exponent of the attractiveness distribution $\gamma_2$. Error-bars are standard deviations obtained by averaging across $10^3$ simulations. We considered $N=10^3$, $m=6$, and $\gamma_1=2$.}
\label{MFPT_1}
\end{figure}

\begin{figure}
\includegraphics[width=0.5\textwidth]{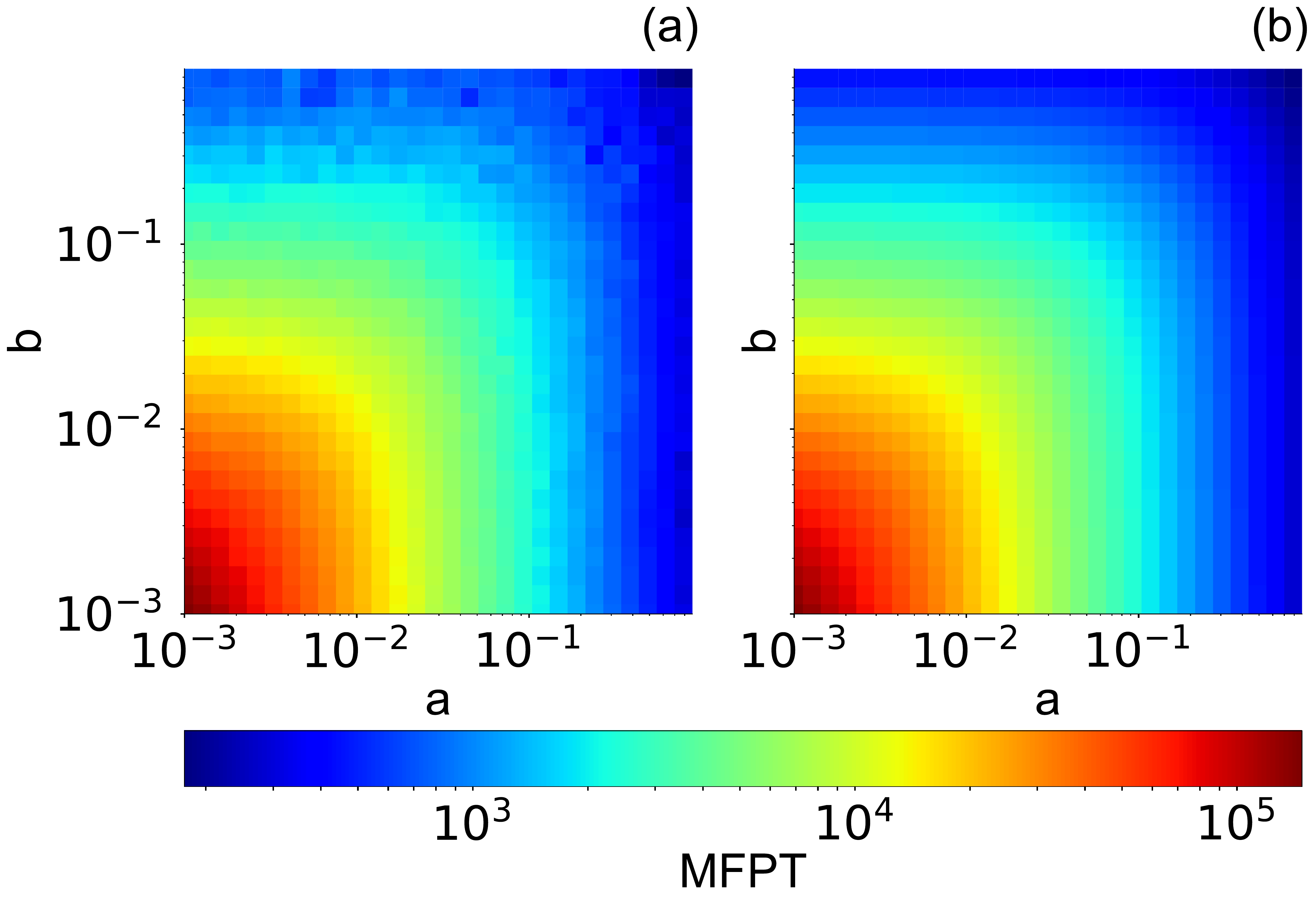}
\caption{\textbf{Mean first passage time for nodes of class $(a,b)$.} Heat map for the average values of MFPT per node of class $(a,b)$ computed through numerical simulations (a) and analytically (b). Colors are attributed based on the value of MFPT as shown in the colorbar on the bottom of the figure. We considered $N=10^3$, $m=6$, $\gamma_1=2$, and $\gamma_2=2$. }
\label{MFPT_2}
\end{figure}

In Figure~\ref{MFPT_1} we test the validity of the analytical expression for the MFPT. We fixed $\gamma_1$ and considered different values of $\gamma_2$ assuming uncorrelated activities and attractiveness. In Figure~\ref{MFPT_2} we show the comparison between the average values of MFPT for nodes of class $a$ and $b$. In both cases we find very good agreement between theory and simulations. It is interesting to observe that the effect of heterogeneous attractiveness is to introduce delays in the transport dynamics since the MFPT is larger for all nodes with respect to the random-tie selection process case (Figure \ref{MFPT_1}, bottom panel, black line)

\section{Conclusions}
\label{conclu}

We presented a model of time-varying networks in which nodes are characterised by activity and attractiveness, regulating their propensity to initiate an interaction and their popularity, respectively. In particular, we extended the framework of activity-driven networks by introducing a tie selection mechanism based on a global linear preferential attachment. We grounded our model with empirical observations by measuring activity and attractiveness from the out-strength and in-strength of nodes in two real time-varying networks describing interactions between i) users on Facebook and ii) people involved in the development of a software. Interestingly, we observed that both activity and attractiveness are heterogeneously distributed and correlated. In the two datasets the correlation is positive.

We then studied the interplay between activity and attractiveness and its effects on the prototypical random walk process. We derived analytical expressions for the stationary state and for the MFPT of the process unfolding on the time-varying network model. We thoroughly tested  the analytical predictions via large-scale numerical simulations obtaining very good agreement between the two. Overall, the results shed light on how the presence of temporal connectivity patterns significantly alters the standard picture obtained in static and annealed networks. The presence of a global tie selection process and the possible correlation between activity and attractiveness introduce non-trivial effects. The stationary state and MFPT are significantly different from those obtained in activity-driven networks characterised by a random tie selection mechanism. In the uncorrelated case the effect of heterogeneous attractiveness is to limit the capability of very active nodes to gather walkers. In the case of positive correlations between activity and attractiveness, observed in real scenarios, the stationary state of the process is substantially altered: The average number of walkers per node decreases as a function of the node activity if activity and attractiveness are different, it is constant if they are the same. Heterogeneous attractiveness furthermore slows down the transport dynamics, as we observe that in this case the MFPT is larger for all nodes.

The presented model can be further enriched in several ways. In particular, the activation dynamics it describes is Poissonian rather than bursty as typically observed in real systems \cite{barabasi2005origin, 0295-5075-81-4-48002,PhysRevE.73.036127, jo2012circadian,Karsai:2012aa, PhysRevE.83.025102, 10.1371/journal.pone.0040612,PhysRevLett.114.108701,ubaldi2016burstiness}. The tie selection process is driven only by global popularity and neglects local tie reinforcement mechanisms responsible for high-order organisation of real networks. The framework of activity-driven networks has been extended in several instances to include such features~\cite{karsai13-1,ubaldi2015asymptotic,ubaldi2016burstiness}. However, the study of nodes' popularity and its effect on networks' dynamical properties was missing. Overall, the results presented in this article contribute towards the development of a comprehensive picture about how the dynamics \emph{of} networks affect the dynamics unfolding \emph{upon} networks.

\bibliography{refs}

\begin{thebibliography}{74}
\expandafter\ifx\csname natexlab\endcsname\relax\def\natexlab#1{#1}\fi
\expandafter\ifx\csname bibnamefont\endcsname\relax
  \def\bibnamefont#1{#1}\fi
\expandafter\ifx\csname bibfnamefont\endcsname\relax
  \def\bibfnamefont#1{#1}\fi
\expandafter\ifx\csname citenamefont\endcsname\relax
  \def\citenamefont#1{#1}\fi
\expandafter\ifx\csname url\endcsname\relax
  \def\url#1{\texttt{#1}}\fi
\expandafter\ifx\csname urlprefix\endcsname\relax\def\urlprefix{URL }\fi
\providecommand{\bibinfo}[2]{#2}
\providecommand{\eprint}[2][]{\url{#2}}

\bibitem[{\citenamefont{Newman}(2010)}]{newman10-1}
\bibinfo{author}{\bibfnamefont{M.}~\bibnamefont{Newman}},
  \emph{\bibinfo{title}{Networks. An Introduction}} (\bibinfo{publisher}{Oxford
  Univesity Press}, \bibinfo{year}{2010}).

\bibitem[{\citenamefont{Jackson et~al.}(2008)}]{jackson2008social}
\bibinfo{author}{\bibfnamefont{M.~O.} \bibnamefont{Jackson}}
  \bibnamefont{et~al.}, \emph{\bibinfo{title}{Social and economic networks}},
  vol.~\bibinfo{volume}{3} (\bibinfo{publisher}{Princeton university press
  Princeton}, \bibinfo{year}{2008}).

\bibitem[{\citenamefont{Gon{\c{c}}alves and
  Perra}(2015)}]{gonccalves2015social}
\bibinfo{author}{\bibfnamefont{B.}~\bibnamefont{Gon{\c{c}}alves}}
  \bibnamefont{and} \bibinfo{author}{\bibfnamefont{N.}~\bibnamefont{Perra}},
  \emph{\bibinfo{title}{Social phenomena: From data analysis to models}}
  (\bibinfo{publisher}{Springer}, \bibinfo{year}{2015}).

\bibitem[{\citenamefont{Holme and Saram\"aki}(2012)}]{holme11-1}
\bibinfo{author}{\bibfnamefont{P.}~\bibnamefont{Holme}} \bibnamefont{and}
  \bibinfo{author}{\bibfnamefont{J.}~\bibnamefont{Saram\"aki}},
  \bibinfo{journal}{Phys. Rep.} \textbf{\bibinfo{volume}{519}},
  \bibinfo{pages}{97} (\bibinfo{year}{2012}).

\bibitem[{\citenamefont{Holme}(2015)}]{holme2015modern}
\bibinfo{author}{\bibfnamefont{P.}~\bibnamefont{Holme}}, \bibinfo{journal}{The
  European Physical Journal B} \textbf{\bibinfo{volume}{88}},
  \bibinfo{pages}{1} (\bibinfo{year}{2015}).

\bibitem[{\citenamefont{Barrat et~al.}(2008)\citenamefont{Barrat,
  Barth{\'e}lemy, and Vespignani}}]{barrat08-1}
\bibinfo{author}{\bibfnamefont{A.}~\bibnamefont{Barrat}},
  \bibinfo{author}{\bibfnamefont{M.}~\bibnamefont{Barth{\'e}lemy}},
  \bibnamefont{and}
  \bibinfo{author}{\bibfnamefont{A.}~\bibnamefont{Vespignani}},
  \emph{\bibinfo{title}{Dynamical Processes on Complex Networks}}
  (\bibinfo{publisher}{Cambridge Univesity Press}, \bibinfo{year}{2008}).

\bibitem[{\citenamefont{Masuda and Lambiotte}(2016)}]{masuda2016guide}
\bibinfo{author}{\bibfnamefont{N.}~\bibnamefont{Masuda}} \bibnamefont{and}
  \bibinfo{author}{\bibfnamefont{R.}~\bibnamefont{Lambiotte}},
  \emph{\bibinfo{title}{A guide to temporal networks}},
  vol.~\bibinfo{volume}{4} (\bibinfo{publisher}{World Scientific},
  \bibinfo{year}{2016}).

\bibitem[{\citenamefont{Perra et~al.}(2012{\natexlab{a}})\citenamefont{Perra,
  Gon\c{c}alves, Pastor-Satorras, and Vespignani}}]{perra12-1}
\bibinfo{author}{\bibfnamefont{N.}~\bibnamefont{Perra}},
  \bibinfo{author}{\bibfnamefont{B.}~\bibnamefont{Gon\c{c}alves}},
  \bibinfo{author}{\bibfnamefont{R.}~\bibnamefont{Pastor-Satorras}},
  \bibnamefont{and}
  \bibinfo{author}{\bibfnamefont{A.}~\bibnamefont{Vespignani}},
  \bibinfo{journal}{Scientific Reports} \textbf{\bibinfo{volume}{2}},
  \bibinfo{pages}{469} (\bibinfo{year}{2012}{\natexlab{a}}).

\bibitem[{\citenamefont{Karsai et~al.}(2014)\citenamefont{Karsai, Perra, and
  Vespignani}}]{karsai13-1}
\bibinfo{author}{\bibfnamefont{M.}~\bibnamefont{Karsai}},
  \bibinfo{author}{\bibfnamefont{N.}~\bibnamefont{Perra}}, \bibnamefont{and}
  \bibinfo{author}{\bibfnamefont{A.}~\bibnamefont{Vespignani}},
  \bibinfo{journal}{Scientific Reports} \textbf{\bibinfo{volume}{4}},
  \bibinfo{pages}{4001} (\bibinfo{year}{2014}).

\bibitem[{\citenamefont{Ubaldi et~al.}(2016)\citenamefont{Ubaldi, Perra,
  Karsai, Vezzani, Burioni, and Vespignani}}]{ubaldi2015asymptotic}
\bibinfo{author}{\bibfnamefont{E.}~\bibnamefont{Ubaldi}},
  \bibinfo{author}{\bibfnamefont{N.}~\bibnamefont{Perra}},
  \bibinfo{author}{\bibfnamefont{M.}~\bibnamefont{Karsai}},
  \bibinfo{author}{\bibfnamefont{A.}~\bibnamefont{Vezzani}},
  \bibinfo{author}{\bibfnamefont{R.}~\bibnamefont{Burioni}}, \bibnamefont{and}
  \bibinfo{author}{\bibfnamefont{A.}~\bibnamefont{Vespignani}},
  \bibinfo{journal}{Scientific Reports} \textbf{\bibinfo{volume}{6}}
  (\bibinfo{year}{2016}).

\bibitem[{\citenamefont{Laurent et~al.}(2015)\citenamefont{Laurent,
  Saram{\"a}ki, and Karsai}}]{laurent2015calls}
\bibinfo{author}{\bibfnamefont{G.}~\bibnamefont{Laurent}},
  \bibinfo{author}{\bibfnamefont{J.}~\bibnamefont{Saram{\"a}ki}},
  \bibnamefont{and} \bibinfo{author}{\bibfnamefont{M.}~\bibnamefont{Karsai}},
  \bibinfo{journal}{The European Physical Journal B}
  \textbf{\bibinfo{volume}{88}}, \bibinfo{pages}{1} (\bibinfo{year}{2015}).

\bibitem[{\citenamefont{Miritello et~al.}(2011)\citenamefont{Miritello, Moro,
  and Lara}}]{miritello2011dynamical}
\bibinfo{author}{\bibfnamefont{G.}~\bibnamefont{Miritello}},
  \bibinfo{author}{\bibfnamefont{E.}~\bibnamefont{Moro}}, \bibnamefont{and}
  \bibinfo{author}{\bibfnamefont{R.}~\bibnamefont{Lara}},
  \bibinfo{journal}{Physical Review E} \textbf{\bibinfo{volume}{83}},
  \bibinfo{pages}{045102} (\bibinfo{year}{2011}),
  \urlprefix\url{http://dx.doi.org/10.1103/PhysRevE.83.045102}.

\bibitem[{\citenamefont{Clauset and Eagle}(2007)}]{clauset07}
\bibinfo{author}{\bibfnamefont{A.}~\bibnamefont{Clauset}} \bibnamefont{and}
  \bibinfo{author}{\bibfnamefont{N.}~\bibnamefont{Eagle}}, in
  \emph{\bibinfo{booktitle}{DIMACS Workshop on Computational Methods for
  Dynamic Interaction Networks}} (\bibinfo{year}{2007}), pp.
  \bibinfo{pages}{1--5}.

\bibitem[{\citenamefont{Isella et~al.}(2011)\citenamefont{Isella, Stehl\'e,
  Barrat, Cattuto, Pinton, and den Broeck}}]{Isella:2011}
\bibinfo{author}{\bibfnamefont{L.}~\bibnamefont{Isella}},
  \bibinfo{author}{\bibfnamefont{J.}~\bibnamefont{Stehl\'e}},
  \bibinfo{author}{\bibfnamefont{A.}~\bibnamefont{Barrat}},
  \bibinfo{author}{\bibfnamefont{C.}~\bibnamefont{Cattuto}},
  \bibinfo{author}{\bibfnamefont{J.-F.} \bibnamefont{Pinton}},
  \bibnamefont{and} \bibinfo{author}{\bibfnamefont{W.~V.} \bibnamefont{den
  Broeck}}, \bibinfo{journal}{J. Theor. Biol} \textbf{\bibinfo{volume}{271}},
  \bibinfo{pages}{166} (\bibinfo{year}{2011}).

\bibitem[{\citenamefont{Saram{\"a}ki and Moro}(2015)}]{saramaki2015seconds}
\bibinfo{author}{\bibfnamefont{J.}~\bibnamefont{Saram{\"a}ki}}
  \bibnamefont{and} \bibinfo{author}{\bibfnamefont{E.}~\bibnamefont{Moro}},
  \bibinfo{journal}{The European Physical Journal B}
  \textbf{\bibinfo{volume}{88}}, \bibinfo{pages}{1} (\bibinfo{year}{2015}).

\bibitem[{\citenamefont{Saram{\"a}ki et~al.}(2014)\citenamefont{Saram{\"a}ki,
  Leicht, L{\'o}pez, Roberts, Reed-Tsochas, and Dunbar}}]{Saramaki21012014}
\bibinfo{author}{\bibfnamefont{J.}~\bibnamefont{Saram{\"a}ki}},
  \bibinfo{author}{\bibfnamefont{E.~A.} \bibnamefont{Leicht}},
  \bibinfo{author}{\bibfnamefont{E.}~\bibnamefont{L{\'o}pez}},
  \bibinfo{author}{\bibfnamefont{S.~G.~B.} \bibnamefont{Roberts}},
  \bibinfo{author}{\bibfnamefont{F.}~\bibnamefont{Reed-Tsochas}},
  \bibnamefont{and} \bibinfo{author}{\bibfnamefont{R.~I.~M.}
  \bibnamefont{Dunbar}}, \bibinfo{journal}{Proceedings of the National Academy
  of Sciences} \textbf{\bibinfo{volume}{111}}, \bibinfo{pages}{942}
  (\bibinfo{year}{2014}),
  \eprint{http://www.pnas.org/content/111/3/942.full.pdf+html},
  \urlprefix\url{http://www.pnas.org/content/111/3/942.abstract}.

\bibitem[{\citenamefont{Sekara et~al.}(2016)\citenamefont{Sekara, Stopczynski,
  and Lehmann}}]{Sekara06092016}
\bibinfo{author}{\bibfnamefont{V.}~\bibnamefont{Sekara}},
  \bibinfo{author}{\bibfnamefont{A.}~\bibnamefont{Stopczynski}},
  \bibnamefont{and} \bibinfo{author}{\bibfnamefont{S.}~\bibnamefont{Lehmann}},
  \bibinfo{journal}{Proceedings of the National Academy of Sciences}
  \textbf{\bibinfo{volume}{113}}, \bibinfo{pages}{9977} (\bibinfo{year}{2016}),
  \eprint{http://www.pnas.org/content/113/36/9977.full.pdf},
  \urlprefix\url{http://www.pnas.org/content/113/36/9977.abstract}.

\bibitem[{\citenamefont{Tomasello et~al.}(2014)\citenamefont{Tomasello, Perra,
  Tessone, Karsai, and Schweitzer}}]{tomasello2014role}
\bibinfo{author}{\bibfnamefont{M.}~\bibnamefont{Tomasello}},
  \bibinfo{author}{\bibfnamefont{N.}~\bibnamefont{Perra}},
  \bibinfo{author}{\bibfnamefont{C.}~\bibnamefont{Tessone}},
  \bibinfo{author}{\bibfnamefont{M.}~\bibnamefont{Karsai}}, \bibnamefont{and}
  \bibinfo{author}{\bibfnamefont{F.}~\bibnamefont{Schweitzer}},
  \bibinfo{journal}{Scientific reports} \textbf{\bibinfo{volume}{4}}
  (\bibinfo{year}{2014}).

\bibitem[{\citenamefont{Barrat and Cattuto}(2015)}]{barrat2015face}
\bibinfo{author}{\bibfnamefont{A.}~\bibnamefont{Barrat}} \bibnamefont{and}
  \bibinfo{author}{\bibfnamefont{C.}~\bibnamefont{Cattuto}}, in
  \emph{\bibinfo{booktitle}{Social Phenomena}} (\bibinfo{publisher}{Springer
  International Publishing}, \bibinfo{year}{2015}), pp.
  \bibinfo{pages}{37--57}.

\bibitem[{\citenamefont{Perra et~al.}(2012{\natexlab{b}})\citenamefont{Perra,
  Baronchelli, Mocanu, Gon\c{c}alves, Pastor-Satorras, and
  Vespignani}}]{perra12-2}
\bibinfo{author}{\bibfnamefont{N.}~\bibnamefont{Perra}},
  \bibinfo{author}{\bibfnamefont{A.}~\bibnamefont{Baronchelli}},
  \bibinfo{author}{\bibfnamefont{D.}~\bibnamefont{Mocanu}},
  \bibinfo{author}{\bibfnamefont{B.}~\bibnamefont{Gon\c{c}alves}},
  \bibinfo{author}{\bibfnamefont{R.}~\bibnamefont{Pastor-Satorras}},
  \bibnamefont{and}
  \bibinfo{author}{\bibfnamefont{A.}~\bibnamefont{Vespignani}},
  \bibinfo{journal}{Phys. Rev. Lett.} \textbf{\bibinfo{volume}{109}},
  \bibinfo{pages}{238701} (\bibinfo{year}{2012}{\natexlab{b}}).

\bibitem[{\citenamefont{Ribeiro et~al.}(2013)\citenamefont{Ribeiro, Perra, and
  Baronchelli}}]{ribeiro12-2}
\bibinfo{author}{\bibfnamefont{B.}~\bibnamefont{Ribeiro}},
  \bibinfo{author}{\bibfnamefont{N.}~\bibnamefont{Perra}}, \bibnamefont{and}
  \bibinfo{author}{\bibfnamefont{A.}~\bibnamefont{Baronchelli}},
  \bibinfo{journal}{Scientific Reports} \textbf{\bibinfo{volume}{3}},
  \bibinfo{pages}{3006} (\bibinfo{year}{2013}).

\bibitem[{\citenamefont{Liu et~al.}(2014)\citenamefont{Liu, Perra, Karsai, and
  Vespignani}}]{PhysRevLett.112.118702}
\bibinfo{author}{\bibfnamefont{S.}~\bibnamefont{Liu}},
  \bibinfo{author}{\bibfnamefont{N.}~\bibnamefont{Perra}},
  \bibinfo{author}{\bibfnamefont{M.}~\bibnamefont{Karsai}}, \bibnamefont{and}
  \bibinfo{author}{\bibfnamefont{A.}~\bibnamefont{Vespignani}},
  \bibinfo{journal}{Phys. Rev. Lett.} \textbf{\bibinfo{volume}{112}},
  \bibinfo{pages}{118702} (\bibinfo{year}{2014}),
  \urlprefix\url{http://link.aps.org/doi/10.1103/PhysRevLett.112.118702}.

\bibitem[{\citenamefont{Liu et~al.}(2013)\citenamefont{Liu, Baronchelli, and
  Perra}}]{PhysRevE.87.032805}
\bibinfo{author}{\bibfnamefont{S.-Y.} \bibnamefont{Liu}},
  \bibinfo{author}{\bibfnamefont{A.}~\bibnamefont{Baronchelli}},
  \bibnamefont{and} \bibinfo{author}{\bibfnamefont{N.}~\bibnamefont{Perra}},
  \bibinfo{journal}{Phys. Rev. E} \textbf{\bibinfo{volume}{87}},
  \bibinfo{pages}{032805} (\bibinfo{year}{2013}),
  \urlprefix\url{http://link.aps.org/doi/10.1103/PhysRevE.87.032805}.

\bibitem[{\citenamefont{Ren and Wang}(2014)}]{10.1063}
\bibinfo{author}{\bibfnamefont{G.}~\bibnamefont{Ren}} \bibnamefont{and}
  \bibinfo{author}{\bibfnamefont{X.}~\bibnamefont{Wang}},
  \bibinfo{journal}{Chaos: An Interdisciplinary Journal of Nonlinear Science}
  \textbf{\bibinfo{volume}{24}}, \bibinfo{eid}{023116} (\bibinfo{year}{2014}),
  \urlprefix\url{http://scitation.aip.org/content/aip/journal/chaos/24/2/10.1063/1.4876436}.

\bibitem[{\citenamefont{Starnini
  et~al.}(2013{\natexlab{a}})\citenamefont{Starnini, Machens, Cattuto, Barrat,
  and Pastor-Satorras}}]{starnini13-1}
\bibinfo{author}{\bibfnamefont{M.}~\bibnamefont{Starnini}},
  \bibinfo{author}{\bibfnamefont{A.}~\bibnamefont{Machens}},
  \bibinfo{author}{\bibfnamefont{C.}~\bibnamefont{Cattuto}},
  \bibinfo{author}{\bibfnamefont{A.}~\bibnamefont{Barrat}}, \bibnamefont{and}
  \bibinfo{author}{\bibfnamefont{R.}~\bibnamefont{Pastor-Satorras}},
  \bibinfo{journal}{Journal of Theoretical Biology}
  \textbf{\bibinfo{volume}{337}}, \bibinfo{pages}{89}
  (\bibinfo{year}{2013}{\natexlab{a}}).

\bibitem[{\citenamefont{Starnini et~al.}(2012)\citenamefont{Starnini,
  Baronchelli, Barrat, and Pastor-Satorras}}]{starnini_rw_temp_nets}
\bibinfo{author}{\bibfnamefont{M.}~\bibnamefont{Starnini}},
  \bibinfo{author}{\bibfnamefont{A.}~\bibnamefont{Baronchelli}},
  \bibinfo{author}{\bibfnamefont{A.}~\bibnamefont{Barrat}}, \bibnamefont{and}
  \bibinfo{author}{\bibfnamefont{R.}~\bibnamefont{Pastor-Satorras}},
  \bibinfo{journal}{Phys. Rev. E} \textbf{\bibinfo{volume}{85}},
  \bibinfo{pages}{056115} (\bibinfo{year}{2012}).

\bibitem[{\citenamefont{Valdano et~al.}(2015)\citenamefont{Valdano, Ferreri,
  Poletto, and Colizza}}]{valdano2015analytical}
\bibinfo{author}{\bibfnamefont{E.}~\bibnamefont{Valdano}},
  \bibinfo{author}{\bibfnamefont{L.}~\bibnamefont{Ferreri}},
  \bibinfo{author}{\bibfnamefont{C.}~\bibnamefont{Poletto}}, \bibnamefont{and}
  \bibinfo{author}{\bibfnamefont{V.}~\bibnamefont{Colizza}},
  \bibinfo{journal}{Physical Review X} \textbf{\bibinfo{volume}{5}},
  \bibinfo{pages}{021005} (\bibinfo{year}{2015}).

\bibitem[{\citenamefont{Scholtes et~al.}(2014)\citenamefont{Scholtes, Wider,
  Pfitzner, Garas, Tessone, and Schweitzer}}]{scholtes2014causality}
\bibinfo{author}{\bibfnamefont{I.}~\bibnamefont{Scholtes}},
  \bibinfo{author}{\bibfnamefont{N.}~\bibnamefont{Wider}},
  \bibinfo{author}{\bibfnamefont{R.}~\bibnamefont{Pfitzner}},
  \bibinfo{author}{\bibfnamefont{A.}~\bibnamefont{Garas}},
  \bibinfo{author}{\bibfnamefont{C.}~\bibnamefont{Tessone}}, \bibnamefont{and}
  \bibinfo{author}{\bibfnamefont{F.}~\bibnamefont{Schweitzer}},
  \bibinfo{journal}{Nature communications} \textbf{\bibinfo{volume}{5}}
  (\bibinfo{year}{2014}).

\bibitem[{\citenamefont{Williams and Musolesi}(2016)}]{Williams160196}
\bibinfo{author}{\bibfnamefont{M.~J.} \bibnamefont{Williams}} \bibnamefont{and}
  \bibinfo{author}{\bibfnamefont{M.}~\bibnamefont{Musolesi}},
  \bibinfo{journal}{Royal Society Open Science} \textbf{\bibinfo{volume}{3}}
  (\bibinfo{year}{2016}).

\bibitem[{\citenamefont{Rocha and Masuda}(2014)}]{rocha2014random}
\bibinfo{author}{\bibfnamefont{L.~E.} \bibnamefont{Rocha}} \bibnamefont{and}
  \bibinfo{author}{\bibfnamefont{N.}~\bibnamefont{Masuda}},
  \bibinfo{journal}{New Journal of Physics} \textbf{\bibinfo{volume}{16}},
  \bibinfo{pages}{063023} (\bibinfo{year}{2014}).

\bibitem[{\citenamefont{Takaguchi
  et~al.}(2012{\natexlab{a}})\citenamefont{Takaguchi, Sato, Yano, and
  Masuda}}]{takaguchi2012importance}
\bibinfo{author}{\bibfnamefont{T.}~\bibnamefont{Takaguchi}},
  \bibinfo{author}{\bibfnamefont{N.}~\bibnamefont{Sato}},
  \bibinfo{author}{\bibfnamefont{K.}~\bibnamefont{Yano}}, \bibnamefont{and}
  \bibinfo{author}{\bibfnamefont{N.}~\bibnamefont{Masuda}},
  \bibinfo{journal}{New Journal of Physics} \textbf{\bibinfo{volume}{14}},
  \bibinfo{pages}{093003} (\bibinfo{year}{2012}{\natexlab{a}}).

\bibitem[{\citenamefont{Rocha and Blondel}(2013)}]{rocha2013bursts}
\bibinfo{author}{\bibfnamefont{L.~E.} \bibnamefont{Rocha}} \bibnamefont{and}
  \bibinfo{author}{\bibfnamefont{V.~D.} \bibnamefont{Blondel}},
  \bibinfo{journal}{PLoS Comput Biol} \textbf{\bibinfo{volume}{9}},
  \bibinfo{pages}{e1002974} (\bibinfo{year}{2013}).

\bibitem[{\citenamefont{Ghoshal and Holme}(2006)}]{ghoshal2006attractiveness}
\bibinfo{author}{\bibfnamefont{G.}~\bibnamefont{Ghoshal}} \bibnamefont{and}
  \bibinfo{author}{\bibfnamefont{P.}~\bibnamefont{Holme}},
  \bibinfo{journal}{Physica A: Statistical Mechanics and its Applications}
  \textbf{\bibinfo{volume}{364}}, \bibinfo{pages}{603} (\bibinfo{year}{2006}).

\bibitem[{\citenamefont{Sun et~al.}(2015)\citenamefont{Sun, Baronchelli, and
  Perra}}]{sun2015contrasting}
\bibinfo{author}{\bibfnamefont{K.}~\bibnamefont{Sun}},
  \bibinfo{author}{\bibfnamefont{A.}~\bibnamefont{Baronchelli}},
  \bibnamefont{and} \bibinfo{author}{\bibfnamefont{N.}~\bibnamefont{Perra}},
  \bibinfo{journal}{The European Physical Journal B}
  \textbf{\bibinfo{volume}{88}}, \bibinfo{pages}{1} (\bibinfo{year}{2015}).

\bibitem[{\citenamefont{Pfitzner et~al.}(2013)\citenamefont{Pfitzner, Scholtes,
  Garas, Tessone, and Schweitzer}}]{pfitzner2013betweenness}
\bibinfo{author}{\bibfnamefont{R.}~\bibnamefont{Pfitzner}},
  \bibinfo{author}{\bibfnamefont{I.}~\bibnamefont{Scholtes}},
  \bibinfo{author}{\bibfnamefont{A.}~\bibnamefont{Garas}},
  \bibinfo{author}{\bibfnamefont{C.~J.} \bibnamefont{Tessone}},
  \bibnamefont{and}
  \bibinfo{author}{\bibfnamefont{F.}~\bibnamefont{Schweitzer}},
  \bibinfo{journal}{Physical review letters} \textbf{\bibinfo{volume}{110}},
  \bibinfo{pages}{198701} (\bibinfo{year}{2013}).

\bibitem[{\citenamefont{Takaguchi
  et~al.}(2012{\natexlab{b}})\citenamefont{Takaguchi, Sato, Yano, and
  Masuda}}]{takaguchi12-1}
\bibinfo{author}{\bibfnamefont{T.}~\bibnamefont{Takaguchi}},
  \bibinfo{author}{\bibfnamefont{N.}~\bibnamefont{Sato}},
  \bibinfo{author}{\bibfnamefont{K.}~\bibnamefont{Yano}}, \bibnamefont{and}
  \bibinfo{author}{\bibfnamefont{N.}~\bibnamefont{Masuda}},
  \bibinfo{journal}{New J. Phys.} \textbf{\bibinfo{volume}{14}},
  \bibinfo{pages}{093003} (\bibinfo{year}{2012}{\natexlab{b}}).

\bibitem[{\citenamefont{Takaguchi et~al.}(2013)\citenamefont{Takaguchi, Masuda,
  and Holme}}]{takaguchi2013bursty}
\bibinfo{author}{\bibfnamefont{T.}~\bibnamefont{Takaguchi}},
  \bibinfo{author}{\bibfnamefont{N.}~\bibnamefont{Masuda}}, \bibnamefont{and}
  \bibinfo{author}{\bibfnamefont{P.}~\bibnamefont{Holme}},
  \bibinfo{journal}{PloS one} \textbf{\bibinfo{volume}{8}},
  \bibinfo{pages}{e68629} (\bibinfo{year}{2013}).

\bibitem[{\citenamefont{Holme and Liljeros}(2014)}]{holme2014birth}
\bibinfo{author}{\bibfnamefont{P.}~\bibnamefont{Holme}} \bibnamefont{and}
  \bibinfo{author}{\bibfnamefont{F.}~\bibnamefont{Liljeros}},
  \bibinfo{journal}{Scientific reports} \textbf{\bibinfo{volume}{4}}
  (\bibinfo{year}{2014}).

\bibitem[{\citenamefont{Holme and Masuda}(2015)}]{holme2015basic}
\bibinfo{author}{\bibfnamefont{P.}~\bibnamefont{Holme}} \bibnamefont{and}
  \bibinfo{author}{\bibfnamefont{N.}~\bibnamefont{Masuda}},
  \bibinfo{journal}{PloS one} \textbf{\bibinfo{volume}{10}},
  \bibinfo{pages}{e0120567} (\bibinfo{year}{2015}).

\bibitem[{\citenamefont{Kivela et~al.}(2012)\citenamefont{Kivela, {Kumar Pan},
  Kaski, Kertesz, Saramaki, and Karsai}}]{dynnetkaski2011}
\bibinfo{author}{\bibfnamefont{M.}~\bibnamefont{Kivela}},
  \bibinfo{author}{\bibfnamefont{R.}~\bibnamefont{{Kumar Pan}}},
  \bibinfo{author}{\bibfnamefont{K.}~\bibnamefont{Kaski}},
  \bibinfo{author}{\bibfnamefont{J.}~\bibnamefont{Kertesz}},
  \bibinfo{author}{\bibfnamefont{J.}~\bibnamefont{Saramaki}}, \bibnamefont{and}
  \bibinfo{author}{\bibfnamefont{M.}~\bibnamefont{Karsai}},
  \bibinfo{journal}{J. Stat. Mech.} \textbf{\bibinfo{volume}{03005}}
  (\bibinfo{year}{2012}).

\bibitem[{\citenamefont{Hoffmann et~al.}(2012)\citenamefont{Hoffmann, Porter,
  and Lambiotte}}]{hoffmann2012generalized}
\bibinfo{author}{\bibfnamefont{T.}~\bibnamefont{Hoffmann}},
  \bibinfo{author}{\bibfnamefont{M.~A.} \bibnamefont{Porter}},
  \bibnamefont{and}
  \bibinfo{author}{\bibfnamefont{R.}~\bibnamefont{Lambiotte}},
  \bibinfo{journal}{Physical Review E} \textbf{\bibinfo{volume}{86}},
  \bibinfo{pages}{046102} (\bibinfo{year}{2012}).

\bibitem[{\citenamefont{Wang et~al.}(2016)\citenamefont{Wang, Bauch,
  Bhattacharyya, d?Onofrio, Manfredi, Perc, Perra, Salath{\'e}, and
  Zhao}}]{wang2016statistical}
\bibinfo{author}{\bibfnamefont{Z.}~\bibnamefont{Wang}},
  \bibinfo{author}{\bibfnamefont{C.~T.} \bibnamefont{Bauch}},
  \bibinfo{author}{\bibfnamefont{S.}~\bibnamefont{Bhattacharyya}},
  \bibinfo{author}{\bibfnamefont{A.}~\bibnamefont{d?Onofrio}},
  \bibinfo{author}{\bibfnamefont{P.}~\bibnamefont{Manfredi}},
  \bibinfo{author}{\bibfnamefont{M.}~\bibnamefont{Perc}},
  \bibinfo{author}{\bibfnamefont{N.}~\bibnamefont{Perra}},
  \bibinfo{author}{\bibfnamefont{M.}~\bibnamefont{Salath{\'e}}},
  \bibnamefont{and} \bibinfo{author}{\bibfnamefont{D.}~\bibnamefont{Zhao}},
  \bibinfo{journal}{Physics Reports}  (\bibinfo{year}{2016}).

\bibitem[{\citenamefont{Fournet and Barrat}(2014)}]{fournet2014contact}
\bibinfo{author}{\bibfnamefont{J.}~\bibnamefont{Fournet}} \bibnamefont{and}
  \bibinfo{author}{\bibfnamefont{A.}~\bibnamefont{Barrat}},
  \bibinfo{journal}{PloS one} \textbf{\bibinfo{volume}{9}},
  \bibinfo{pages}{e107878} (\bibinfo{year}{2014}).

\bibitem[{\citenamefont{Karsai et~al.}(2011{\natexlab{a}})\citenamefont{Karsai,
  Kivel\"{a}, Pan, Kaski, Kert\'{e}sz, Barab\'{a}si, and
  Saram\"{a}ki}}]{Karsai2011Small}
\bibinfo{author}{\bibfnamefont{M.}~\bibnamefont{Karsai}},
  \bibinfo{author}{\bibfnamefont{M.}~\bibnamefont{Kivel\"{a}}},
  \bibinfo{author}{\bibfnamefont{R.~K.} \bibnamefont{Pan}},
  \bibinfo{author}{\bibfnamefont{K.}~\bibnamefont{Kaski}},
  \bibinfo{author}{\bibfnamefont{J.}~\bibnamefont{Kert\'{e}sz}},
  \bibinfo{author}{\bibfnamefont{A.~L.} \bibnamefont{Barab\'{a}si}},
  \bibnamefont{and}
  \bibinfo{author}{\bibfnamefont{J.}~\bibnamefont{Saram\"{a}ki}},
  \bibinfo{journal}{Physical Review E} \textbf{\bibinfo{volume}{83}},
  \bibinfo{pages}{025102} (\bibinfo{year}{2011}{\natexlab{a}}),
  \urlprefix\url{http://dx.doi.org/10.1103/PhysRevE.83.025102}.

\bibitem[{\citenamefont{Moinet et~al.}(2015{\natexlab{a}})\citenamefont{Moinet,
  Starnini, and Pastor-Satorras}}]{moinet2015burstiness}
\bibinfo{author}{\bibfnamefont{A.}~\bibnamefont{Moinet}},
  \bibinfo{author}{\bibfnamefont{M.}~\bibnamefont{Starnini}}, \bibnamefont{and}
  \bibinfo{author}{\bibfnamefont{R.}~\bibnamefont{Pastor-Satorras}},
  \bibinfo{journal}{Physical review letters} \textbf{\bibinfo{volume}{114}},
  \bibinfo{pages}{108701} (\bibinfo{year}{2015}{\natexlab{a}}).

\bibitem[{\citenamefont{Karsai et~al.}(2012{\natexlab{a}})\citenamefont{Karsai,
  Kaski, Barab\'asi, and Kert\'esz}}]{karsai12-1}
\bibinfo{author}{\bibfnamefont{M.}~\bibnamefont{Karsai}},
  \bibinfo{author}{\bibfnamefont{K.}~\bibnamefont{Kaski}},
  \bibinfo{author}{\bibfnamefont{A.-L.} \bibnamefont{Barab\'asi}},
  \bibnamefont{and}
  \bibinfo{author}{\bibfnamefont{J.}~\bibnamefont{Kert\'esz}},
  \bibinfo{journal}{Scientific Reports} p. \bibinfo{pages}{397}
  (\bibinfo{year}{2012}{\natexlab{a}}).

\bibitem[{\citenamefont{Ubaldi et~al.}(2017)\citenamefont{Ubaldi, Vezzani,
  Karsai, Perra, and Burioni}}]{ubaldi2016burstiness}
\bibinfo{author}{\bibfnamefont{E.}~\bibnamefont{Ubaldi}},
  \bibinfo{author}{\bibfnamefont{A.}~\bibnamefont{Vezzani}},
  \bibinfo{author}{\bibfnamefont{M.}~\bibnamefont{Karsai}},
  \bibinfo{author}{\bibfnamefont{N.}~\bibnamefont{Perra}}, \bibnamefont{and}
  \bibinfo{author}{\bibfnamefont{R.}~\bibnamefont{Burioni}},
  \bibinfo{journal}{Scientific Reports} \textbf{\bibinfo{volume}{7}}
  (\bibinfo{year}{2017}).

\bibitem[{\citenamefont{Onnela et~al.}(2007)\citenamefont{Onnela, Saram\"aki,
  Hyv\"onen, Szab\'o, Lazer, Kaski, Kert\'esz, and Barab\'asi}}]{Onnela:2007}
\bibinfo{author}{\bibfnamefont{J.-P.} \bibnamefont{Onnela}},
  \bibinfo{author}{\bibfnamefont{J.}~\bibnamefont{Saram\"aki}},
  \bibinfo{author}{\bibfnamefont{J.}~\bibnamefont{Hyv\"onen}},
  \bibinfo{author}{\bibfnamefont{G.}~\bibnamefont{Szab\'o}},
  \bibinfo{author}{\bibfnamefont{D.}~\bibnamefont{Lazer}},
  \bibinfo{author}{\bibfnamefont{K.}~\bibnamefont{Kaski}},
  \bibinfo{author}{\bibfnamefont{J.}~\bibnamefont{Kert\'esz}},
  \bibnamefont{and} \bibinfo{author}{\bibfnamefont{A.-L.}
  \bibnamefont{Barab\'asi}}, \bibinfo{journal}{Proceedings of the National
  Academy of Sciences} \textbf{\bibinfo{volume}{104}}, \bibinfo{pages}{7332}
  (\bibinfo{year}{2007}),
  \urlprefix\url{http://www.pnas.org/content/104/18/7332.abstract}.

\bibitem[{\citenamefont{Granovetter}(1973)}]{granovetter73-1}
\bibinfo{author}{\bibfnamefont{M.}~\bibnamefont{Granovetter}},
  \bibinfo{journal}{Am. J. Sociol.} \textbf{\bibinfo{volume}{78}},
  \bibinfo{pages}{1360} (\bibinfo{year}{1973}).

\bibitem[{\citenamefont{Starnini and
  Pastor-Satorras}(2013)}]{PhysRevE.87.062807}
\bibinfo{author}{\bibfnamefont{M.}~\bibnamefont{Starnini}} \bibnamefont{and}
  \bibinfo{author}{\bibfnamefont{R.}~\bibnamefont{Pastor-Satorras}},
  \bibinfo{journal}{Phys. Rev. E} \textbf{\bibinfo{volume}{87}},
  \bibinfo{pages}{062807} (\bibinfo{year}{2013}),
  \urlprefix\url{https://link.aps.org/doi/10.1103/PhysRevE.87.062807}.

\bibitem[{\citenamefont{Starnini
  et~al.}(2013{\natexlab{b}})\citenamefont{Starnini, Baronchelli, and
  Pastor-Satorras}}]{starnini2013modeling}
\bibinfo{author}{\bibfnamefont{M.}~\bibnamefont{Starnini}},
  \bibinfo{author}{\bibfnamefont{A.}~\bibnamefont{Baronchelli}},
  \bibnamefont{and}
  \bibinfo{author}{\bibfnamefont{R.}~\bibnamefont{Pastor-Satorras}},
  \bibinfo{journal}{Physical Review Letters} \textbf{\bibinfo{volume}{110}},
  \bibinfo{pages}{168701} (\bibinfo{year}{2013}{\natexlab{b}}).

\bibitem[{\citenamefont{Starnini
  et~al.}(2016{\natexlab{a}})\citenamefont{Starnini, Baronchelli, and
  Pastor-Satorras}}]{starnini2016model}
\bibinfo{author}{\bibfnamefont{M.}~\bibnamefont{Starnini}},
  \bibinfo{author}{\bibfnamefont{A.}~\bibnamefont{Baronchelli}},
  \bibnamefont{and}
  \bibinfo{author}{\bibfnamefont{R.}~\bibnamefont{Pastor-Satorras}},
  \bibinfo{journal}{Social Networks} \textbf{\bibinfo{volume}{47}},
  \bibinfo{pages}{130} (\bibinfo{year}{2016}{\natexlab{a}}).

\bibitem[{\citenamefont{Mariani et~al.}(2015)\citenamefont{Mariani, Medo, and
  Zhang}}]{mariani2015ranking}
\bibinfo{author}{\bibfnamefont{M.~S.} \bibnamefont{Mariani}},
  \bibinfo{author}{\bibfnamefont{M.}~\bibnamefont{Medo}}, \bibnamefont{and}
  \bibinfo{author}{\bibfnamefont{Y.-C.} \bibnamefont{Zhang}},
  \bibinfo{journal}{Scientific reports} \textbf{\bibinfo{volume}{5}}
  (\bibinfo{year}{2015}).

\bibitem[{\citenamefont{Barabasi}(2016)}]{barabasi2016network}
\bibinfo{author}{\bibfnamefont{A.-L.} \bibnamefont{Barabasi}},
  \emph{\bibinfo{title}{Network science}} (\bibinfo{publisher}{Cambridge
  University Press}, \bibinfo{year}{2016}).

\bibitem[{\citenamefont{Lambiotte et~al.}(2013)\citenamefont{Lambiotte,
  Tabourier, and Delvenne}}]{lambiotte2013burstiness}
\bibinfo{author}{\bibfnamefont{R.}~\bibnamefont{Lambiotte}},
  \bibinfo{author}{\bibfnamefont{L.}~\bibnamefont{Tabourier}},
  \bibnamefont{and} \bibinfo{author}{\bibfnamefont{J.-C.}
  \bibnamefont{Delvenne}}, \bibinfo{journal}{The European Physical Journal B}
  \textbf{\bibinfo{volume}{86}}, \bibinfo{pages}{1} (\bibinfo{year}{2013}).

\bibitem[{\citenamefont{Lambiotte et~al.}(2015)\citenamefont{Lambiotte,
  Salnikov, and Rosvall}}]{lambiotte2015effect}
\bibinfo{author}{\bibfnamefont{R.}~\bibnamefont{Lambiotte}},
  \bibinfo{author}{\bibfnamefont{V.}~\bibnamefont{Salnikov}}, \bibnamefont{and}
  \bibinfo{author}{\bibfnamefont{M.}~\bibnamefont{Rosvall}},
  \bibinfo{journal}{Journal of Complex Networks} \textbf{\bibinfo{volume}{3}},
  \bibinfo{pages}{177} (\bibinfo{year}{2015}).

\bibitem[{\citenamefont{Starnini
  et~al.}(2016{\natexlab{b}})\citenamefont{Starnini, Frasca, and
  Baronchelli}}]{starnini2016emergence}
\bibinfo{author}{\bibfnamefont{M.}~\bibnamefont{Starnini}},
  \bibinfo{author}{\bibfnamefont{M.}~\bibnamefont{Frasca}}, \bibnamefont{and}
  \bibinfo{author}{\bibfnamefont{A.}~\bibnamefont{Baronchelli}},
  \bibinfo{journal}{Scientific Reports} \textbf{\bibinfo{volume}{6}},
  \bibinfo{pages}{31834} (\bibinfo{year}{2016}{\natexlab{b}}).

\bibitem[{\citenamefont{Gillespie}(1977)}]{gillespie1977exact}
\bibinfo{author}{\bibfnamefont{D.~T.} \bibnamefont{Gillespie}},
  \bibinfo{journal}{The journal of physical chemistry}
  \textbf{\bibinfo{volume}{81}}, \bibinfo{pages}{2340} (\bibinfo{year}{1977}).

\bibitem[{\citenamefont{Viswanath et~al.}(2009)\citenamefont{Viswanath,
  Mislove, Cha, and Gummadi}}]{viswanath2009evolution}
\bibinfo{author}{\bibfnamefont{B.}~\bibnamefont{Viswanath}},
  \bibinfo{author}{\bibfnamefont{A.}~\bibnamefont{Mislove}},
  \bibinfo{author}{\bibfnamefont{M.}~\bibnamefont{Cha}}, \bibnamefont{and}
  \bibinfo{author}{\bibfnamefont{K.~P.} \bibnamefont{Gummadi}}, in
  \emph{\bibinfo{booktitle}{Proceedings of the 2nd ACM workshop on Online
  social networks}} (\bibinfo{organization}{ACM}, \bibinfo{year}{2009}), pp.
  \bibinfo{pages}{37--42}.

\bibitem[{kon(2016{\natexlab{a}})}]{konect:2016:facebook-wosn-wall}
\emph{\bibinfo{title}{Facebook wall posts network dataset -- {KONECT}}}
  (\bibinfo{year}{2016}{\natexlab{a}}),
  \urlprefix\url{http://konect.uni-koblenz.de/networks/facebook-wosn-wall}.

\bibitem[{kon(2016{\natexlab{b}})}]{konect:2016:lkml-reply}
\emph{\bibinfo{title}{Linux kernel mailing list replies network dataset --
  {KONECT}}} (\bibinfo{year}{2016}{\natexlab{b}}),
  \urlprefix\url{http://konect.uni-koblenz.de/networks/lkml-reply}.

\bibitem[{\citenamefont{Barrat et~al.}(2004)\citenamefont{Barrat, Barthelemy,
  Pastor-Satorras, and Vespignani}}]{barrat2004architecture}
\bibinfo{author}{\bibfnamefont{A.}~\bibnamefont{Barrat}},
  \bibinfo{author}{\bibfnamefont{M.}~\bibnamefont{Barthelemy}},
  \bibinfo{author}{\bibfnamefont{R.}~\bibnamefont{Pastor-Satorras}},
  \bibnamefont{and}
  \bibinfo{author}{\bibfnamefont{A.}~\bibnamefont{Vespignani}},
  \bibinfo{journal}{Proceedings of the National Academy of Sciences of the
  United States of America} \textbf{\bibinfo{volume}{101}},
  \bibinfo{pages}{3747} (\bibinfo{year}{2004}).

\bibitem[{\citenamefont{Noh and Rieger}(2004)}]{noh04}
\bibinfo{author}{\bibfnamefont{J.}~\bibnamefont{Noh}} \bibnamefont{and}
  \bibinfo{author}{\bibfnamefont{H.}~\bibnamefont{Rieger}},
  \bibinfo{journal}{Phys. Rev. Lett.} \textbf{\bibinfo{volume}{92}},
  \bibinfo{pages}{118701} (\bibinfo{year}{2004}).

\bibitem[{\citenamefont{Vespignani}(2012)}]{alex12-1}
\bibinfo{author}{\bibfnamefont{A.}~\bibnamefont{Vespignani}},
  \bibinfo{journal}{Nature Physics} \textbf{\bibinfo{volume}{8}},
  \bibinfo{pages}{32} (\bibinfo{year}{2012}).

\bibitem[{\citenamefont{Redner}(2001)}]{Redner01}
\bibinfo{author}{\bibfnamefont{S.}~\bibnamefont{Redner}},
  \emph{\bibinfo{title}{A Guide To First-Passage Processes}}
  (\bibinfo{publisher}{Cambridge University Press},
  \bibinfo{address}{Cambridge}, \bibinfo{year}{2001}).

\bibitem[{\citenamefont{Baronchelli and Loreto}(2006)}]{baronchelli2006ring}
\bibinfo{author}{\bibfnamefont{A.}~\bibnamefont{Baronchelli}} \bibnamefont{and}
  \bibinfo{author}{\bibfnamefont{V.}~\bibnamefont{Loreto}},
  \bibinfo{journal}{Physical Review E} \textbf{\bibinfo{volume}{73}},
  \bibinfo{pages}{026103} (\bibinfo{year}{2006}).

\bibitem[{\citenamefont{Barabasi}(2005)}]{barabasi2005origin}
\bibinfo{author}{\bibfnamefont{A.-L.} \bibnamefont{Barabasi}},
  \bibinfo{journal}{Nature} \textbf{\bibinfo{volume}{435}},
  \bibinfo{pages}{207} (\bibinfo{year}{2005}).

\bibitem[{\citenamefont{Goh and Barab{\'a}si}(2008)}]{0295-5075-81-4-48002}
\bibinfo{author}{\bibfnamefont{K.-I.} \bibnamefont{Goh}} \bibnamefont{and}
  \bibinfo{author}{\bibfnamefont{A.-L.} \bibnamefont{Barab{\'a}si}},
  \bibinfo{journal}{EPL (Europhysics Letters)} \textbf{\bibinfo{volume}{81}},
  \bibinfo{pages}{48002} (\bibinfo{year}{2008}),
  \urlprefix\url{http://stacks.iop.org/0295-5075/81/i=4/a=48002}.

\bibitem[{\citenamefont{V\'azquez et~al.}(2006)\citenamefont{V\'azquez,
  Oliveira, Dezs\"o, Goh, Kondor, and Barab\'asi}}]{PhysRevE.73.036127}
\bibinfo{author}{\bibfnamefont{A.}~\bibnamefont{V\'azquez}},
  \bibinfo{author}{\bibfnamefont{J.~a.~G.} \bibnamefont{Oliveira}},
  \bibinfo{author}{\bibfnamefont{Z.}~\bibnamefont{Dezs\"o}},
  \bibinfo{author}{\bibfnamefont{K.-I.} \bibnamefont{Goh}},
  \bibinfo{author}{\bibfnamefont{I.}~\bibnamefont{Kondor}}, \bibnamefont{and}
  \bibinfo{author}{\bibfnamefont{A.-L.} \bibnamefont{Barab\'asi}},
  \bibinfo{journal}{Phys. Rev. E} \textbf{\bibinfo{volume}{73}},
  \bibinfo{pages}{036127} (\bibinfo{year}{2006}),
  \urlprefix\url{http://link.aps.org/doi/10.1103/PhysRevE.73.036127}.

\bibitem[{\citenamefont{Jo et~al.}(2012)\citenamefont{Jo, Karsai, Kert{\'e}sz,
  and Kaski}}]{jo2012circadian}
\bibinfo{author}{\bibfnamefont{H.-H.} \bibnamefont{Jo}},
  \bibinfo{author}{\bibfnamefont{M.}~\bibnamefont{Karsai}},
  \bibinfo{author}{\bibfnamefont{J.}~\bibnamefont{Kert{\'e}sz}},
  \bibnamefont{and} \bibinfo{author}{\bibfnamefont{K.}~\bibnamefont{Kaski}},
  \bibinfo{journal}{New Journal of Physics} \textbf{\bibinfo{volume}{14}},
  \bibinfo{pages}{013055} (\bibinfo{year}{2012}).

\bibitem[{\citenamefont{Karsai et~al.}(2012{\natexlab{b}})\citenamefont{Karsai,
  Kaski, Barab{\'a}si, and Kert{\'e}sz}}]{Karsai:2012aa}
\bibinfo{author}{\bibfnamefont{M.}~\bibnamefont{Karsai}},
  \bibinfo{author}{\bibfnamefont{K.}~\bibnamefont{Kaski}},
  \bibinfo{author}{\bibfnamefont{A.-L.} \bibnamefont{Barab{\'a}si}},
  \bibnamefont{and}
  \bibinfo{author}{\bibfnamefont{J.}~\bibnamefont{Kert{\'e}sz}},
  \bibinfo{journal}{Sci. Rep.} \textbf{\bibinfo{volume}{2}}
  (\bibinfo{year}{2012}{\natexlab{b}}),
  \urlprefix\url{http://dx.doi.org/10.1038/srep00397}.

\bibitem[{\citenamefont{Karsai et~al.}(2011{\natexlab{b}})\citenamefont{Karsai,
  Kivel\"a, Pan, Kaski, Kert\'esz, Barab\'asi, and
  Saram\"aki}}]{PhysRevE.83.025102}
\bibinfo{author}{\bibfnamefont{M.}~\bibnamefont{Karsai}},
  \bibinfo{author}{\bibfnamefont{M.}~\bibnamefont{Kivel\"a}},
  \bibinfo{author}{\bibfnamefont{R.~K.} \bibnamefont{Pan}},
  \bibinfo{author}{\bibfnamefont{K.}~\bibnamefont{Kaski}},
  \bibinfo{author}{\bibfnamefont{J.}~\bibnamefont{Kert\'esz}},
  \bibinfo{author}{\bibfnamefont{A.-L.} \bibnamefont{Barab\'asi}},
  \bibnamefont{and}
  \bibinfo{author}{\bibfnamefont{J.}~\bibnamefont{Saram\"aki}},
  \bibinfo{journal}{Phys. Rev. E} \textbf{\bibinfo{volume}{83}},
  \bibinfo{pages}{025102} (\bibinfo{year}{2011}{\natexlab{b}}),
  \urlprefix\url{http://link.aps.org/doi/10.1103/PhysRevE.83.025102}.

\bibitem[{\citenamefont{Karsai et~al.}(2012{\natexlab{c}})\citenamefont{Karsai,
  Kaski, and Kert{\'e}sz}}]{10.1371/journal.pone.0040612}
\bibinfo{author}{\bibfnamefont{M.}~\bibnamefont{Karsai}},
  \bibinfo{author}{\bibfnamefont{K.}~\bibnamefont{Kaski}}, \bibnamefont{and}
  \bibinfo{author}{\bibfnamefont{J.}~\bibnamefont{Kert{\'e}sz}},
  \bibinfo{journal}{PLoS ONE} \textbf{\bibinfo{volume}{7}},
  \bibinfo{pages}{e40612} (\bibinfo{year}{2012}{\natexlab{c}}),
  \urlprefix\url{http://dx.doi.org/10.1371%2Fjournal.pone.0040612}.

\bibitem[{\citenamefont{Moinet et~al.}(2015{\natexlab{b}})\citenamefont{Moinet,
  Starnini, and Pastor-Satorras}}]{PhysRevLett.114.108701}
\bibinfo{author}{\bibfnamefont{A.}~\bibnamefont{Moinet}},
  \bibinfo{author}{\bibfnamefont{M.}~\bibnamefont{Starnini}}, \bibnamefont{and}
  \bibinfo{author}{\bibfnamefont{R.}~\bibnamefont{Pastor-Satorras}},
  \bibinfo{journal}{Phys. Rev. Lett.} \textbf{\bibinfo{volume}{114}},
  \bibinfo{pages}{108701} (\bibinfo{year}{2015}{\natexlab{b}}),
  \urlprefix\url{http://link.aps.org/doi/10.1103/PhysRevLett.114.108701}.

\end{thebibliography}
\end{document}